 \documentclass[aps, twocolumn, showpacs, amsmath]{revtex4-2}

\usepackage{bm}
\usepackage{graphicx}
\usepackage{color}
\usepackage{hyperref}
\usepackage{amsthm}
\usepackage{amssymb}
\usepackage[sort&compress]{natbib}
\usepackage{dsfont}
\usepackage{footmisc}

\begin{document}

\def\be{\begin{equation}}
\def\ee{\end{equation}}

\def\bc{\begin{center}}
\def\ec{\end{center}}
\def\bea{\begin{eqnarray}}
\def\eea{\end{eqnarray}}
\newcommand{\avg}[1]{\langle{#1}\rangle}
\newcommand{\Avg}[1]{\left\langle{#1}\right\rangle}

\def\ie{\textit{i.e.}}
\def\etal{\textit{et al.}}
\def\m{\vec{m}}
\def\G{\mathcal{G}}

\newcommand{\davide}[1]{{\bf\color{blue}#1}}
\newcommand{\gin}[1]{{\bf\color{green}#1}}

\title{Dimensionality-induced dynamical phase transition in the large deviation  of local time density for Brownian motion}
\author{Ruofei Yan}
\author{Hanshuang Chen}\email{chenhshf@ahu.edu.cn}
\affiliation{School of Physics and Optoelectronic Engineering, Anhui University, Hefei 230601, China}
\begin{abstract}
We study the fluctuation properties of the local time density, ${\rho _T} = \frac{1}{T}\int_0^T {\delta ( {r(t) - 1} )} dt$, spent by a $d$-dimensional Brownian particle at a spherical shell of unit radius, where $r(t)$ denotes the radial distance from the particle to the origin. In the large observation time limit, $T \to \infty$, the local time density $\rho_T$ obeys the large deviation principle, $P(\rho _T= \rho) \sim e^{-T I(\rho)}$, where the rate function $I(\rho)$ is analytic everywhere for $d\leq 4$. In contrast, for $d>4$, $I(\rho)$ becomes nonanalytic at a specific point $\rho=\rho_c^{(d)}$, where $\rho_c^{(d)}=d(d-4)/(2d-4)$ depends solely on dimensionality. The singularity signals the occurrence of a first-order dynamical phase transition in dimensions higher than four. Such a transition is accompanied by temporal phase separations in the large deviations of Brownian trajectories. Finally, we validate our theoretical results using a rare-event simulation approach.   
\end{abstract}

\maketitle

\section{Introduction}\label{sec1}
The study of fluctuations in stochastic systems occupies a central position in nonequilibrium statistical mechanics and probability theory \cite{derrida2007non,bertini2015macroscopic}, with applications spanning nonequilibrium phase transitions \cite{cohen2012phase}, current statistics \cite{bodineau2004current,harris2005current}, population dynamics \cite{ovaskainen2010stochastic,assaf2017wkb}, condensation phenomena \cite{merhav2010bose,szavits2014constraint,smith2022condensation}, fluctuation relations \cite{seifert2012stochastic,barato2015thermodynamic,horowitz2020thermodynamic}, and anomalous scaling \cite{smith2022anomalous,nickelsen2022noise,smith2022anomalous,meerson2019anomalous}. Notably, large deviations—also referred to as rare events—have emerged as a pivotal area of research over the past few decades \cite{ellis2012entropy,oono1989large,hollander2000large,jack2020ergodicity,touchette2009large}. Fluctuations in quantities of interest, such as time-integrated observables, are inherently encoded in large deviation functions (LDFs), which serve as analogs to thermodynamic potentials in equilibrium systems \cite{touchette2009large}.

Among the most remarkable phenomena within the framework of large deviations are dynamical phase transitions (DPTs), defined by the presence of singularities—specifically nonanalyticities—in the LDFs. These transitions have been observed in systems with many degrees of freedom, including lattice-gas models \cite{bodineau2005distribution,bertini2005current,bertini2006non,agranov2023tricritical}, driven diffusive systems \cite{baek2015singularities,baek2017dynamical,bertini2010lagrangian,bunin2013cusp,espigares2013dynamical,aminov2014singularities,shpielberg2017numerical}, kinetically constrained
models of glasses \cite{garrahan2007dynamical,garrahan2009first}, self-propelled particles \cite{kumar2011symmetry,cagnetta2017large,nemoto2019optimizing,semeraro2023work}, random graphs \cite{de2016rare,coghi2019large,carugno2023delocalization}. In systems with only a few degrees of freedom, DPTs have been identified in the weak-noise limit of stochastic dynamics \cite{baek2015singularities,speck2012large,tsobgni2016large,meerson2019geometrical,smith2022anomalous,mukherjee2025nonequilibrium}. More recently, DPTs have also been reported in simpler models that do not require macroscopic or low-noise limits, such as stochastic resetting systems \cite{majumdar2015dynamical,harris2017phase}, drifted Brownian motion \cite{nyawo2017minimal,nyawo2018dynamical} or run-and-tumble motion \cite{mukherjee2024large}, vicious Brownian motions \cite{mukherjee2023dynamical}, nonergodic stochastic processes \cite{yerrababu2024dynamical}, and switching diffusion \cite{gueneau2025large}.

In two recent studies \cite{kanazawa2025universality,kanazawa2024dynamical}, Kanazawa \textit{et al.} demonstrated that DPTs can occur even for a single driftless Brownian particle. They investigated the occupation time statistics of a high-dimensional Brownian particle inside a ball and found that a first-order DPT emerges when the system's dimensionality exceeds four. In the present work, we aim to determine whether such a dimension-induced DPT arises at the large deviation function of another time-integrated quantity: the local time spent by the particle in the neighborhood of a given region during an observation time $T$. This is a crucial quantity with applications across scientific fields, such as in chemical and biological reactions—where reaction rates depend on reactants' local time near receptors \cite{wilemski1973general,doi1975theory,benichou2014first}—and in polymer science (relating to monomer concentration) and bacterial chemotaxis (influenced by bacteria's local time at a point) \cite{koshland1980bacterial}.
The statistics of local time have been extensively studied in contexts such as Ornstein-Uhlenbeck processes \cite{kishore2021local}, diffusion in a random
potential \cite{majumdar2002local,sabhapandit2006statistical}, resetting systems \cite{pal2019local,singh2022first}, diffusion on graphs \cite{comtet2002local}, run-and-tumble particle \cite{singh2021local,mukherjee2024large}, and multi-particle systems \cite{burenev2023local,smith2024macroscopic}.

In this paper, we report a notable new example of dimensionality-induced DPTs in a purely diffusive system. We consider Brownian motion in high-dimensional space, and study the large deviation properties of local time density of the Brownian particle at a spherical shell in the long observation time limit. The LDF $I(\rho)$ exhibits a singularity at $\rho=\rho_c^{(d)}$ when the dimensionality $d$ of the system exceeds a critical dimension $d_c=4$, where $\rho_c^{(d)}$ can be obtained analytically. This singularity indicates that the system undergoes a first-order DPT at $\rho=\rho_c^{(d)}$. The LDF possesses a linear branch for $\rho<\rho_c^{(d)}$ with a slope $k_c^{(d)}=d-2$, and its asymptotics for $\rho>\rho_c^{(d)}$ is deduced. We also demonstrate that such a DPT gives rise to temporal phase separation in the ensemble of dynamical trajectories. Finally, we adopt a statistical-mechanics-inspired sampling method to simulate the rare fluctuations in local time density, and compare the results with the LDFs predicted by our theory, revealing excellent agreement between them.

\section{Model and dynamical observables}\label{sec2}
Let us consider the Brownian motion of a particle in $d$-dimensional space, described by the following Langevin equation,
\begin{eqnarray}\label{eq1.1}
\frac{{d{\mathbf{x}(t)}}}{{dt}} = {\mathbf{\xi }}(t),
\end{eqnarray} 
where ${\bf{x}}(t)=\left(x_1(t),\cdots, x_d(t) \right)^{\top} $ denotes the position of the particle at time $t$, and $\mathbf{\xi }(t)=\left(\xi_1(t),\cdots, \xi_d(t) \right)^{\top} $ is a $d$-dimensional Gaussioan white noise satifying $\langle \xi_i(t) \rangle=0 $ and $\langle \xi_i(t) \xi_j(t') \rangle=2D\delta_{ij}\delta(t-t')$. Without loss of generality, the diffusion constant $D$ is set to unity. 

We focus on the local time density that the particle spends on a $d$-dimensional spherical shell of radius $r_c$. More specifically, we choose $r_c=1$, such that the local time density is defined as 
\begin{eqnarray}\label{eq1.2}
{\rho _T} = \frac{1}{T}\int_0^T {\delta ( {r(t) - 1} )} dt,
\end{eqnarray}
where $r(t)=\left\| {\mathbf{x}(t)} \right\| $ is the distance of the particle from the origin at time $t$. 

For $T \to \infty$, ${\rho _T}$ converges in probability to zero. For large but finite $T$, the probability density of ${\rho _T}$, $P(\rho _T= \rho)$, obeys a large-deviation principle \cite{ellis2012entropy,oono1989large,hollander2000large,jack2020ergodicity,touchette2009large}, 
\begin{eqnarray}\label{eq1.5}
P(\rho _T= \rho)=e^{-T I(\rho) +o(T)},
\end{eqnarray}
with $o(T)/T \to 0$ as $T \to \infty$. Here, $I(\rho)$ is the rate function of local time density. According to the G\"artner-Ellis theorem \cite{ellis2012entropy,oono1989large,hollander2000large,jack2020ergodicity,touchette2009large}, the rate function can be derived via the Legendre-Fenchel transformation of the scaled cumulant generating function (SCGF) $\lambda(k)$,
\begin{eqnarray}\label{eq1.6}
I(\rho)=\mathop {\sup }\limits_k \left[ k \rho- \lambda(k) \right], 
\end{eqnarray}
where the SCGF $\lambda(k)$ is defined as
\begin{eqnarray}\label{eq1.7}
\lambda(k)=\mathop {\lim }\limits_{T \to \infty} \frac{1}{T} \ln \langle e^{Tk \rho_T}\rangle.
\end{eqnarray}
Since probability measures are normalized, $\lambda(0) = 0$. Furthermore,
\begin{eqnarray}\label{eq1.8}
\lambda'(0)=\mathop {\lim }\limits_{T \to \infty}  \langle \rho_T\rangle=0.
\end{eqnarray}
This result follows from a key property of Brownian motion: over long times ($T \to \infty$), it is extremely unlikely for the Brownian particle to remain confined within any finite interval. As a consequence, $\rho_T \to 0$ almost surely (with probability 1) in the long-time limit, which implies the ensemble average $\langle \rho_T\rangle$ vanishes as $T \to \infty$. Additionally, the SCGF $\lambda(k)$ is always convex, which implies it is continuous in the interior of its domain and differentiable everywhere except possibly at a denumerable set of points \cite{touchette2009large}.

For the $d$-dimensional Brownian motion with the time-averaged observable, the SCGF $\lambda(k)$ corresponds to the dominant eigenvalue of the tilted generator \cite{touchette2018introduction},
\begin{eqnarray}\label{eq2.1}
\mathcal{L}_k=\nabla^2 + k \delta(r-1),
\end{eqnarray}
where $\mathbf{\nabla}:=\partial / \partial \mathbf{x}$. The operator $\mathcal{L}_k$ and its eigenvalue $\lambda(k)$ are analogous to the quantum Hamiltonian and the negative of the ground-state energy, respectively, for a particle
in $d$-dimensional space subject to a delta potential $-k \delta (r-1)$. The dominant eigenfunction $\phi_k(\mathbf{x})$ satisfies
\begin{eqnarray}\label{eq2.2}
\mathcal{L}_k \phi_k(\mathbf{x})= \lambda(k) \phi_k(\mathbf{x}).
\end{eqnarray}

\section{Theoretical results}\label{sec3}
The system under study exhibits spherical symmetry. This is because we are seeking the dominant eigenvalue of the operator $\mathcal{L}_k$, which corresponds to the ground state of the quantum Hamiltonian $-\mathcal{L}_k$. In quantum mechanics, it is well-knwon that the ground state maintains radially symmetric when the underlying potential possesses  spherical symmetry. Consequently, the dominant eigenfunction depends solely on the radial coordinate $r=||\mathbf{x}||$, i.e., $\phi_k(\mathbf{x})=\phi_k(r)$. With this simplification,Eq.(\ref{eq2.2}) can be reduced to \cite{kanazawa2025universality}
\begin{eqnarray}\label{eq2.3}
\frac{{{d^2}{\phi _k}( r )}}{{d{r^2}}} + \frac{{d - 1}}{r}\frac{{d{\phi _k}( r )}}{{dr}} + k\delta ( {r - 1} ){\phi _k}( r ) = \lambda ( k ){\phi _k}( r ) . \nonumber \\
\end{eqnarray}
Eq.(\ref{eq2.3}) can be solved for $r<1$ and $r>1$, separately, which yields \cite{arfken2011mathematical}
\begin{eqnarray}\label{eq2.4}
{\phi _k}( r ) = \left\{ \begin{array}{llc}
{A_k}\frac{{{I_{d/2 - 1}}\left( {r\sqrt {\lambda \left( k \right)} } \right)}}{{{{\left( {r\sqrt {\lambda \left( k \right)} } \right)}^{d/2 - 1}}}}, &     0< r< 1,  \\
{B_k}\frac{{{K_{d/2 - 1}}\left( {r\sqrt {\lambda \left( k \right)} } \right)}}{{{{\left( {r\sqrt {\lambda \left( k \right)} } \right)}^{d/2 - 1}}}}, &    r>1.  \\ 
\end{array}  \right.
\end{eqnarray}
Here, $A_k$ and $B_k$ will be determined by matching conditions at $r=1$, and $I_{\nu}(z)$ and $K_{\nu}(z)$ are the modified Bessel functions of first kind and second kind, respectively, with the order $\nu$. The determine the form of the solution, we have imposed physical boundary conditions at $r=0$ and $r \to \infty$: the eigenfunction $\phi_k(r)$ must be regular (finite and well-behaved) at $r=0$, and bounded as $r \to \infty$. Since $K_{\nu}(z)$ diverges at $r=0$ and $I_{\nu}(z)$ does not remain bounded as $r \to \infty$, we select $I_{\nu}(z)$ for $0<r<1$ and $K_{\nu}(z)$ for $r>1$, respectively.  Firstly, the eigenfunction $\phi_k(r)$ is continuous at $r=1$, $\phi_k(1^+)=\phi_k(1^-)$. Secondly, the derivative of $\phi_k(r)$ with respect to $r$ is discontinuous at $r=1$, where the discontinuity can be obtained by integrating Eq.(\ref{eq2.3}) over $r$ from $r-\epsilon$ to $r+\epsilon$ and then letting $\epsilon \to 0^+$. Finally, this procedure leads to $\phi'_k(1^+)-\phi'_k(1^-)=-k \phi_k(1)$. Applying these two matching conditions results in (see Appendix \ref{appa} for the details)
\begin{eqnarray}\label{eq2.5}
\left\{ \begin{gathered}
{A_k}{I_{d/2 - 1}}( {\sqrt \lambda  } ) - {B_k}{K_{d/2 - 1}}( {\sqrt \lambda  } ) = 0, \hfill \\
{A_k}[ {k{I_{d/2 - 1}}( {\sqrt \lambda  } ) - \sqrt \lambda  {I_{d/2}}( {\sqrt \lambda  } )} ] - \sqrt \lambda  {K_{d/2}}( {\sqrt \lambda  } ){B_k} = 0. \hfill \\ 
\end{gathered}  \right. \nonumber \\
\end{eqnarray}
A nontrivial solution to Eq.(\ref{eq2.5}) exists if and only if the determinant of the coefficient matrix vanishes, i.e., 
\begin{eqnarray}\label{eq2.6}
k =\Psi_d(\lambda ),
\end{eqnarray}
with
\begin{eqnarray}\label{eq2.6.1}
\Psi_d(\lambda )= \frac{{\sqrt \lambda  {I_{d/2}}( {\sqrt \lambda  } )}}{{{I_{d/2 - 1}}( {\sqrt \lambda  } )}} + \frac{{\sqrt \lambda  {K_{d/2}}( {\sqrt \lambda  } )}}{{{K_{d/2 - 1}}( {\sqrt \lambda  } )}}.
\end{eqnarray}

Note that Eq.(\ref{eq2.6}) serves as the eigenvalue equation, forming the starting point for calculating the SCGF. If $\lambda(k)>0$, the eigenfunction is localized around $r=r_c$ (analogy to a bound state in quantum mechanics). However, we shall see that the quantum solution obtained from Eq.(\ref{eq2.6}) does not fully 
represent the SCGF because it does not necessarily satisfy $\lambda(0)=0$ \cite{nyawo2017minimal,nyawo2018dynamical,kanazawa2025universality}. To address this, we define a threshold $k_c^{(d)}$ at which the localized state begins to appear. To determine $k_c^{(d)}$, let us analyze the behavior of Eq.(\ref{eq2.6}) in the limit of $\lambda \to 0^+$. For $d\leq 2$, $k \to 0$ as $\lambda \to 0^+$, such that $k_c^{(d)}=0$ for $d \leq 2$. For $d>2$, however, $k \to k_c^{(d)}$ as $\lambda \to 0^+$, where $k_c^{(d)}$ is a nonzero value. To determine $k_c^{(d)}$ for $d>2$, we use the asymptotic forms of modified Bessel functions for small arguments \cite{arfken2011mathematical}, $I_{\nu}(z) \sim \frac{1}{\Gamma(\nu+1)} \left( \frac{z}{2} \right)^{\nu}$ and $K_{\nu}(z) \sim \frac{\Gamma(\nu)}{2}\left( \frac{2}{z} \right)^{\nu} $ for $0<|z| \ll \sqrt{\nu+1}$ and $\nu>0$, to Eq.(\ref{eq2.6}), which leads to 
\begin{eqnarray}\label{eq2.7.1}
k_c^{(d)}=d-2, \quad {\rm{for}} \quad d>2  .
\end{eqnarray}
However, the SCGF $\lambda(k)$ must be nonnegative, as it is convex, $\lambda(0)=0$, and $\lambda'(0)=\lim_{T \to \infty} \langle \rho_T \rangle=0$. Therefore, the SCGF $\lambda(k)$ for $d>2$ is determined by Eq.(\ref{eq2.6}) only when $k>k_c^{(d)}$. While for $k<k_c^{(d)}$, $\lambda(k) = 0$ emerges as another possible eigenvalue of $\mathcal{L}_k$, referred as the ``non-quantum" eigenvalue \cite{nyawo2017minimal,nyawo2018dynamical,kanazawa2025universality}. This implies that there exists a singularity in $\lambda(k)$ at $k=k_c^{(d)}$, a signature of a dynamical phase transition.

To calculate the rate function, we invoke the Legendre-Fenchel transform  (\ref{eq1.6}) of the SCGF $\lambda(k)$. For $k>k_c^{(d)}$, $\lambda(k)$ is everywhere differentiable, so the transform reduces to the better known Legendre transform \cite{touchette2005legendre}: $I(\rho)=k(\rho) \rho-\lambda( k(\rho) )$, where $k(\rho)$ is the unique root of $\lambda '( k )=\rho$. To proceed, we differentiate Eq.(\ref{eq2.6}) with respect to $k$,
\begin{eqnarray}\label{eq2.7.2}
\rho  = \lambda '( k ) = \frac{1}{\Psi'_d(\lambda)},
\end{eqnarray}
with
\begin{eqnarray}\label{eq2.7.3}
\Psi'_d(\lambda)=\frac{d \Psi_d(\lambda)}{d \lambda}=\frac{{{K_{d/2 - 2}}( {\sqrt \lambda  } ){K_{d/2}}( {\sqrt \lambda  } )}}{{2{{\left[ {{K_{d/2 - 1}}( {\sqrt \lambda  } )} \right]}^2}}} \nonumber \\ - \frac{{{}_0{F_1}\left( {d/2 - 1,\lambda /4} \right){}_0{F_1}( {d/2 + 1,\lambda /4} )}}{{2{{\left[ {{}_0{F_1}( {d/2,\lambda /4} )} \right]}^2}}},
\end{eqnarray}
where ${}_0{F_1}(a,z)$ is regularized hypergeometric function. Treating $\lambda$ as a parameter, we use Eq.(\ref{eq2.6}) and Eq.(\ref{eq2.7.2}) to obtain the SCGF and  the rate function, $I(\rho)=k \rho-\lambda( k )$.

\subsection{$d \leq 2$}

\begin{figure}
	\centerline{\includegraphics*[width=1.0\columnwidth]{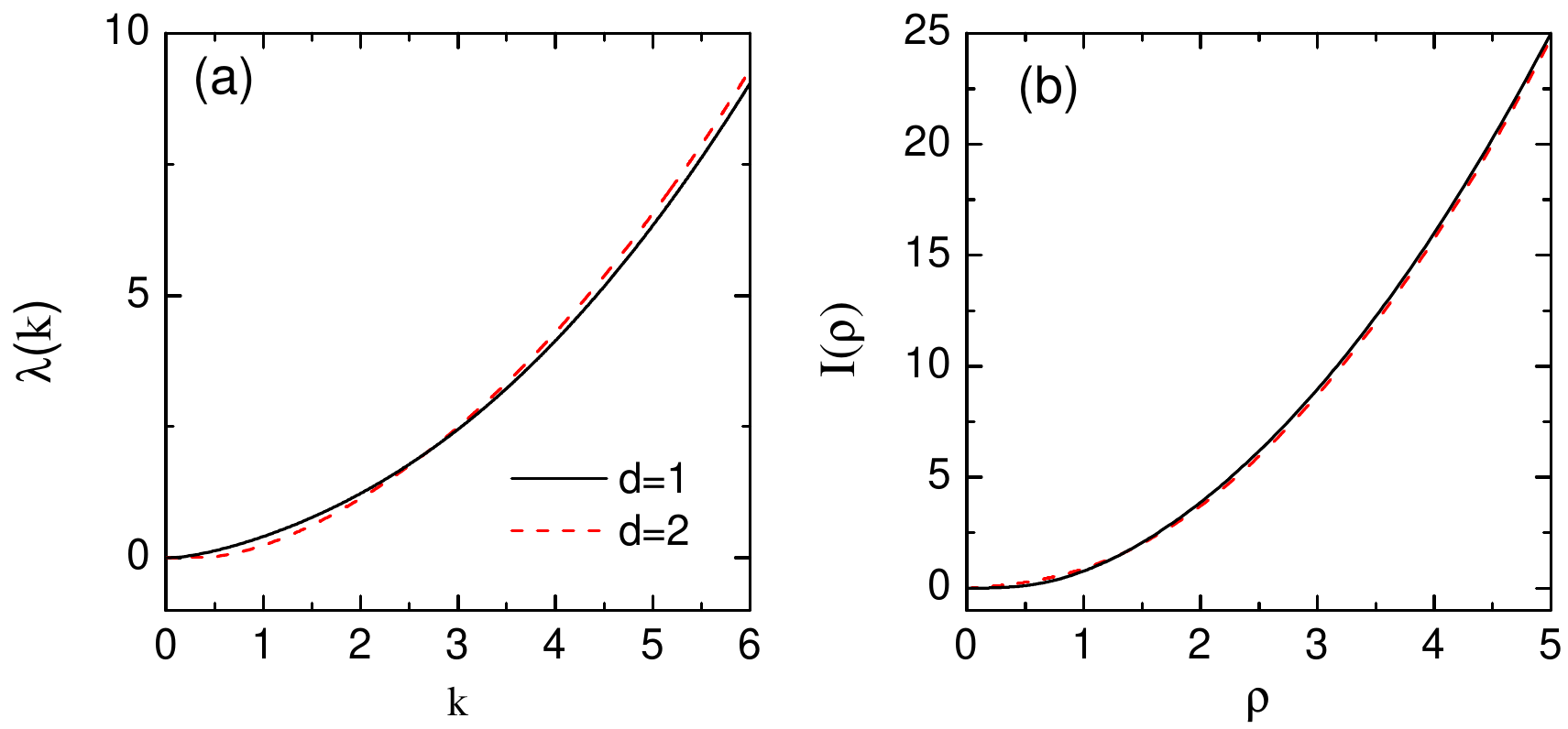}}
	\caption{The results for $d=1$ and $d=2$: The SCGF $\lambda(k)$ (a) and the rate function $I(\rho)$ (b) of local time density $\rho$.    \label{figd1}}
\end{figure}

For $d \leq 2$, the SCGF $\lambda(k)$ is continuous and differentiable everywhere, and thus there is no singularities in $\lambda(k)$ and its Legendre-Fenchel transform $I (\rho)$. This indicates the absence of dynamical phase transitions for $d \leq 2$.

For $d=1$, the modified Bessel functions can be rewritten as the triangle functions \cite{arfken2011mathematical}, $I_{1/2}(z)=\sqrt{\frac{2}{\pi z}} \sinh(z)$, $I_{-1/2}(z)=\sqrt{\frac{2}{\pi z}} \cosh(z)$, $K_{1/2}(z)=K_{-1/2}(z)=\sqrt{\frac{\pi}{2z}}e^{-z}$, and thus Eq.(\ref{eq2.6}) and Eq.(\ref{eq2.7.2}) become
\begin{eqnarray}\label{eq3.1}
k=\left[ 1+\tanh(\sqrt{\lambda})\right] \sqrt{\lambda},
\end{eqnarray}
and
\begin{eqnarray}\label{eq3.2}
\rho  = \lambda '( k ) = \frac{{2\sqrt \lambda  }}{{1 + \sqrt \lambda  {\rm{sech}}^2( {\sqrt \lambda  } ) + \tanh ( {\sqrt \lambda  } )}}.
\end{eqnarray}
For small $k$, $k \sim 0$, Eq.(\ref{eq3.1}) approximates to $\lambda \sim k^2$, and the Legendre-Fenchel transformation gives $I(\rho) \sim \rho^2/4$. 

For $d=2$, Eq.(\ref{eq2.6}) and Eq.(\ref{eq2.7.2}) simplify to
\begin{eqnarray}\label{eq3.3}
k= \frac{{\sqrt \lambda  {I_{1}}( {\sqrt \lambda  } )}}{{{I_{0}}( {\sqrt \lambda  } )}} + \frac{{\sqrt \lambda  {K_{1}}( {\sqrt \lambda  } )}}{{{K_{0}}( {\sqrt \lambda  } )}},
\end{eqnarray}
and 
\begin{eqnarray}\label{eq3.4}
\rho  = \lambda '( k ) = 2{\left\{ {{{\left[ {\frac{{{K_1}( {\sqrt \lambda  } )}}{{{K_0}( {\sqrt \lambda  } )}}} \right]}^2} - {{\left[ {\frac{{{I_1}( {\sqrt \lambda  } )}}{{{I_0}( {\sqrt \lambda  } )}}} \right]}^2}} \right\}^{ - 1}}.
\end{eqnarray}
For $k \sim 0$ and $\lambda \sim 0$, we use the asymptotic expansions of $I_{\nu}(z) \sim \frac{1}{\Gamma(\nu+1)} \left( \frac{z}{2} \right)^{\nu}$, $K_0(z) \sim -\ln(z/2)-\gamma$ and $K_{1}(z) \sim  \frac{1}{z}  $ for small arguments $z$ \cite{arfken2011mathematical}, to Eq.(\ref{eq3.3}), which leads to $\lambda(k) \sim e^{-\frac{1}{k}}$ for $k \ll 1$. Taking the derivative with respect to $k$, one has $\rho =\lambda'(k) \sim \frac{1}{k^2} e^{-\frac{1}{k}}$, from which one obtains ${k}( \rho  ) =-{[ {2{W_{ - 1}}( { - \sqrt {\rho /4} } )} ]^{ - 1}}$. Here, $W_{-1}(z)$ is Lambert $W$ function on branch $-1$. Using the asymptotic form of $W_{-1}(z)$ for $z \ll 1$, $W_{-1}(-z) \sim \ln {z}-\ln(-\ln {z})$, one obtains the asymptotics of the rate function for $\rho \ll 1$,  $I( \rho  ) = {k}( \rho  )\rho  - \lambda \left( {{k}( \rho  )} \right) \sim -\frac{\rho}{\ln{\rho}}$.

In Appendix \ref{appb}, we show that the asymptotic analyses of  $\lambda(k)$ and $I(\rho)$ at large arguments. The results exhibit dimension-independent quadratic scalings $\lambda(k) \sim k^2/4$ ($k \to \infty$), and $I(\rho) \sim \rho^2$ ($\rho \to \infty$), reflecting Gaussian-like fluctuation properties.

In Fig.\ref{figd1}, we show the SCGFs $\lambda(k)$ and the rate function $I( \rho  )$ for $d=1$ and $d=2$. For both dimensions, the SCGF and the rate function are analytic, i.e., they contain no singularities for $k>0$ and $\rho> 0$.

\subsection{$2<d \leq 4$}

For $2<d \leq 4$, the SCGF $\lambda(k)$ is differentiable, as is the rate function $I(\rho)$, meaning no qualitative change in the system’s dynamical paths is expected. However, the second derivative of  $\lambda(k)$ is not continuous at $k= k_c^{(d)}$, which implies that a second-order DPT occurs at $2<d \leq 4$.  We should emphasize that that the DPT classification is based on the singularity order of $\lambda(k)$. To see this, we will discuss the case when $d=3$ and the case when $d=4$, separately.

For $d=3$, we expand Eq.(\ref{eq2.6}) as a series of $\lambda$,
\begin{eqnarray}\label{eq4.1}
k = \Psi_3(\lambda)=1 + \sqrt \lambda   + \frac{\lambda }{3} - \frac{{{\lambda ^2}}}{{45}} + o( {{\lambda ^{5/2}}} ).
\end{eqnarray}
Obviously, $k \to k_c^{(3)}=1$ as $\lambda \to 0^+$. Differentiating Eq.(\ref{eq4.1}) with respect to $k$ and then taking the limit $\lambda \to 0^+$, we have
\begin{subequations}\label{eq4.2}
	\begin{align}
      \lambda '( {k_c^{(3)} +0^+ } ) &= \frac{1}{\Psi'_3(0^+)}=0, \\ \lambda ''({k_c^{(3)} +0^+ } ) &=-\frac{{\Psi''_3(0^+)}}{ [ {\Psi'_3(0^+)} ] ^3  }=2  .
	\end{align}
\end{subequations} 

\begin{figure}
	\centerline{\includegraphics*[width=1.0\columnwidth]{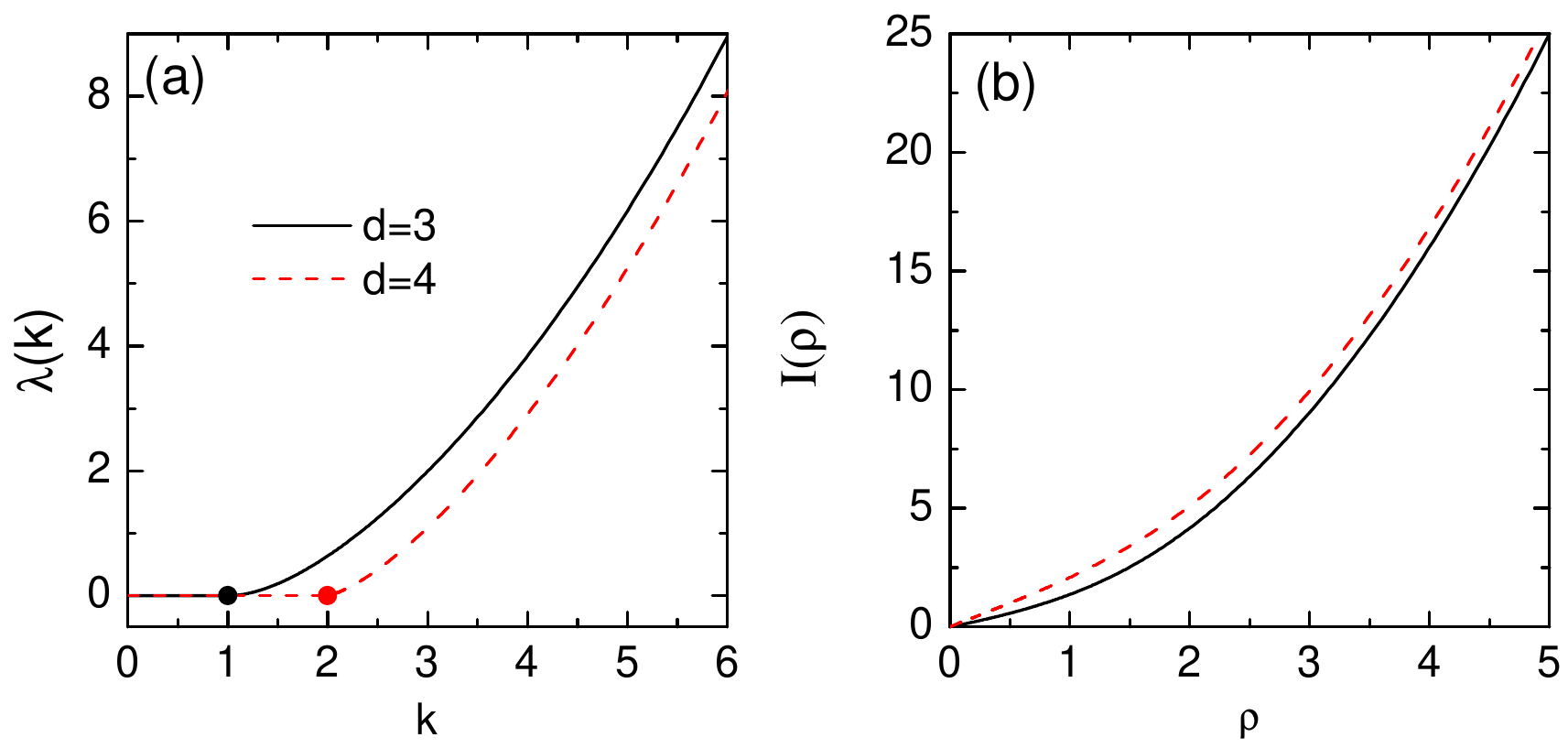}}
	\caption{The results for $d=3$ and $d=4$: The SCGF $\lambda(k)$ (a) and the rate function $I(\rho)$ (b) of local time density $\rho$. The solid circles in (a) denote the point $k=k_c^{(d)}$ where $\lambda(k)$ shows a singularity at this point.   \label{figd3}}
\end{figure}

For $d=4$, we use the same procedure as before, leading to
\begin{eqnarray}\label{eq4.3}
k = \Psi_4(\lambda)=  2 + \frac{\lambda }{4} - \frac{{{\lambda ^2}}}{{96}} - \frac{1}{2}\lambda \ln \lambda + o( {{\lambda ^{5/2}}} ),
\end{eqnarray}
and
\begin{subequations}\label{eq4.4}
	\begin{align}
     \lambda '( {k_c^{(4)} +0^+ } ) &= \frac{1}{\Psi'_4(0^+)}=0, \\ \lambda ''({k_c^{(4)} +0^+ } ) &=-\frac{{\Psi''_4(0^+)}}{ [ {\Psi'_4(0^+)} ] ^3  }=\infty  .
	\end{align}
\end{subequations}

From Eqs.(\ref{eq4.2}) and (\ref{eq4.4}), we find that the first derivative of $\lambda(k)$ is continuous at $k=k_c^{(d)}$ for $d=3$ and $d=4$. However, the second derivative of $\lambda(k)$ is discontinuous at $k=k_c^{(d)}$. 

As shown in Fig.\ref{figd3}, the SCGFs and the rate functions for $d=3$ and $d=4$ are plotted. It can be clearly seen that the SCGF shows a singularity at $k=k_c^{(d)}=d-2$. However, due to the first derivative of $\lambda(k)$ is continuous for all $k>0$, the rate function is analytic, i.e., the rate function contains no singularities.

\subsection{$d>4$}

For $d>4$, the first derivative of the SCGF $\lambda(k)$ is discontinuous at $k=k_c^{(d)}$, which leads to a singularity of the rate function $I(\rho)$ at some point $\rho=\rho_c^{(d)}$. This implies that that the first-order DPTs take place at $d>4$, which is essentially distinct from the second-order singularity observed at $2 < d \leq 4$.  Notably, we will see that the rate functions $I(\rho)$ for $d>4$ have a strictly linear branch at a region $0<\rho<\rho_c^{(d)}$.

\begin{figure}
	\centerline{\includegraphics*[width=1.0\columnwidth]{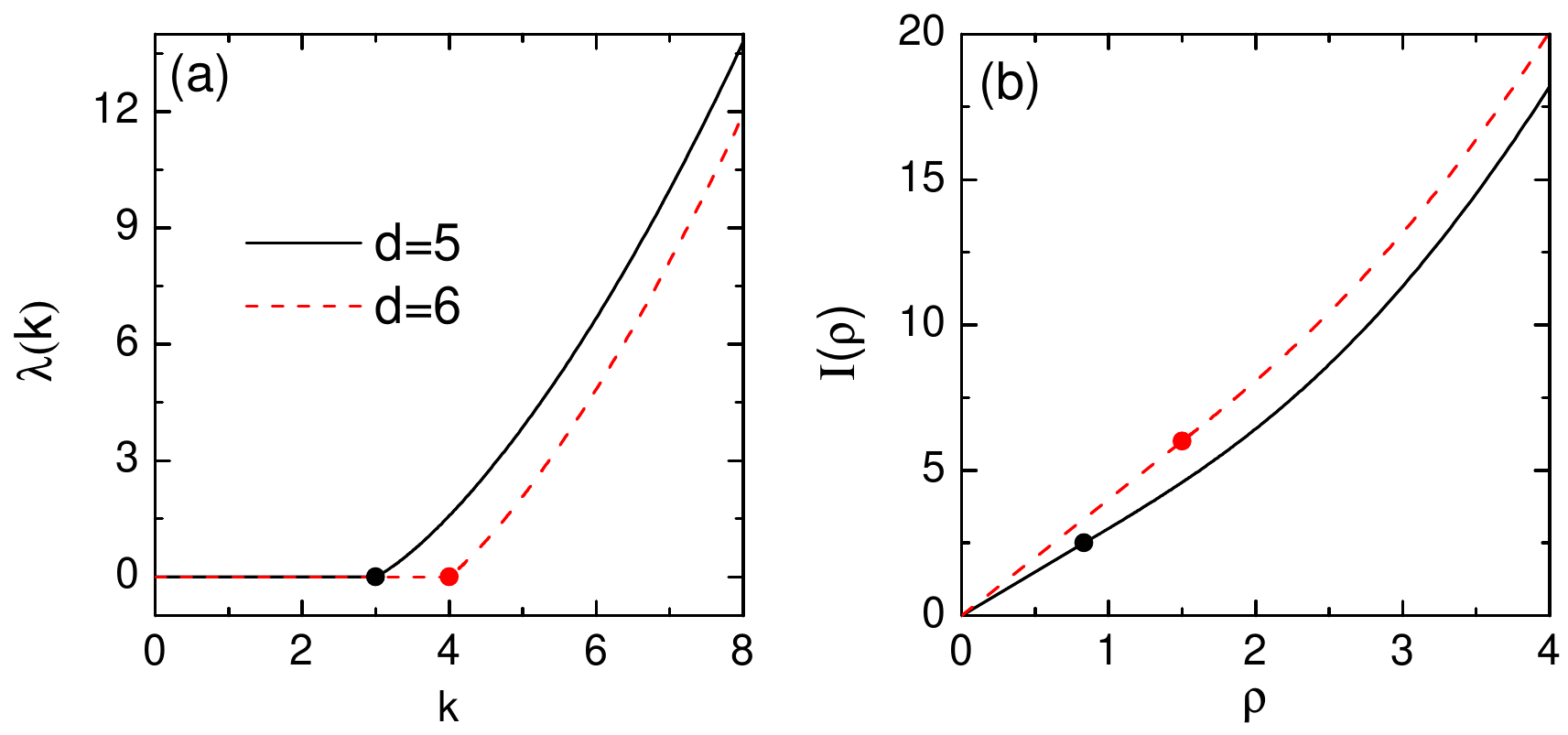}}
	\caption{The results for $d=5$ and $d=6$: The SCGF $\lambda(k)$ (a) and the rate function $I(\rho)$ (b) of local time density $\rho$. The solid circles in (a) denote the point $k=k_c^{(d)}$, at which the first derivative of $\lambda(k)$ is discontinuous. The solid circles in (b) mark the phase transition point $\rho _c^{(d)}=\lambda'(k_c^{(d)}+0^{+})$ below which $I(\rho)$ is linear with the slope $k_c^{(d)}$.   \label{figd5}}
\end{figure}

To determine the first derivative of $\lambda(k)$ just above $k=k_c^{(d)}$. We use the asymptotics of modified Bessel functions to expand Eq.(\ref{eq2.6}) in the limit of $\lambda \to 0^+$, which yields
\begin{eqnarray}\label{eq5.1}
k = \Psi_d(\lambda)=d-2+\frac{2d-4}{d(d-4)}\lambda+o( \lambda  ), \quad {\rm{for}} \quad d>4, \nonumber \\
\end{eqnarray}
where we have used the recursive relation $  {K_{d/2}}( z ) = {K_{d/2-2}}( z )+ \frac{{d-2 }}{z}{K_{d/2-1} }( z )$ \cite{arfken2011mathematical}.
Differentiating Eq.(\ref{eq5.1}) with respect to $k$ and then taking the limit $\lambda \to 0^+$ gives
\begin{eqnarray}\label{eq5.2}
\rho_c^{(d)}&=&\lambda '( {k_c^{(d)} +0^+ } ) = \frac{1}{\Psi'_d(0^+)} \nonumber \\ &=&\frac{{d( {d - 4} )}}{{2d - 4}}, \quad {\rm{for}} \quad d>4 .
\end{eqnarray}

The SCGF $\lambda(k)$ is nondifferentiable at $k=k_c^{(d)}$, as the first derivative of $\lambda(k)$ at $k=k_c^{(d)}$ jumps from $\lambda'(k_c^{(d)}+0^-)=0$ to  $\lambda'(k_c^{(d)}+0^+)=d(d-4)/(2d-4)$, which is nonzero for $d>4$. This yields, by the Legendre-Fenchel transform \cite{nyawo2017minimal}, that the rate functions for $d>4$ have a linear branch at $0<\rho<\rho_c^{(d)}$ with a dimensionality-dependent slope $k_c^{(d)}=d-2$, i.e.,
\begin{eqnarray}\label{eq5.3}
I(\rho)=k_c^{(d)} \rho=(d-2) \rho \quad {\rm{for}} \quad 0<\rho<\rho_c^{(d)}. 
\end{eqnarray}

\begin{figure*}
	\centerline{\includegraphics[width=1.9\columnwidth]{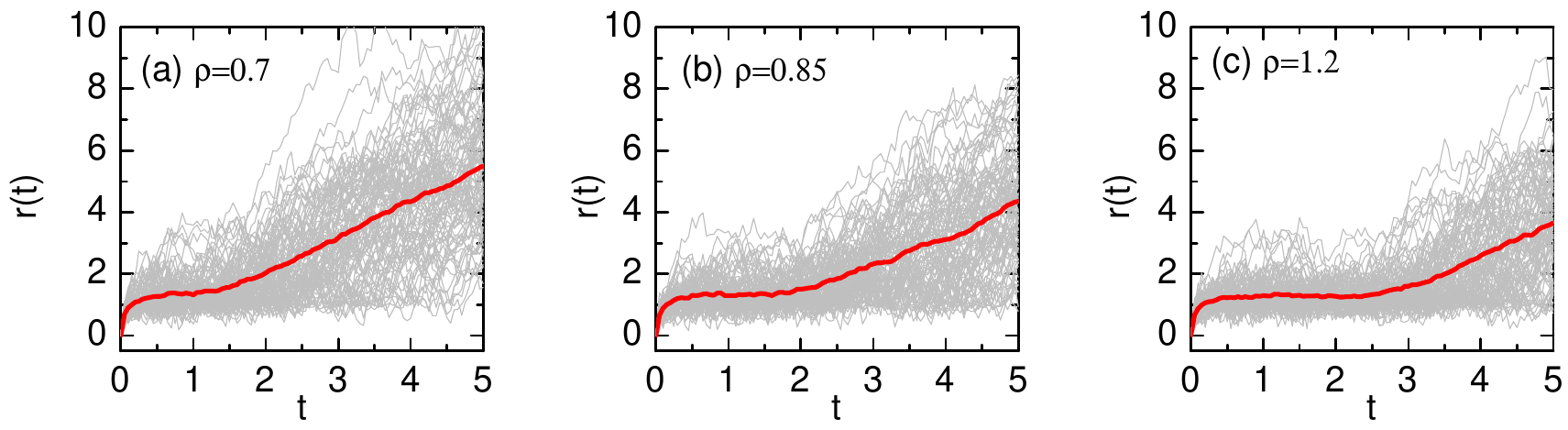}}
	\caption{Sample 100 radial trajectories of the Brownian motion for $d=5$ (grey lines). For each trajectory, the local time density $\rho$ is constrained to be a constant: $\rho=0.7<\rho _c^{(5)}$ (a), $\rho=0.85 \approx \rho _c^{(5)}$ (b), and $\rho=1.2>\rho _c^{(5)}$ (c). The red lines indicate the average values of 100 samples. Parameters: $T=5$, $\Delta t=0.05$, and $\epsilon =0.1$.     \label{figsp}}
\end{figure*}

To obtain the asymptotic form of $I(\rho)$ for $0<\rho-\rho_c^{(d)} \ll 1$, we first expand
Eq.(\ref{eq2.6}) around $\lambda \sim 0$ to obtain, 
\begin{widetext}
\begin{eqnarray}\label{eq5.4}
\delta=k-k_c^{(d)} = \left\{ \begin{array}{llc}
 \frac{6}{5}\lambda  - {\lambda ^{3/2}} + \frac{{174}}{{175}}{\lambda ^2} + o({\lambda ^{5/2}}), &     d=5,  \\
 \frac{2}{3}\lambda  - \frac{{{\lambda ^2}}}{{288}} + \frac{1}{8}{\lambda ^2}\ln \lambda  + \frac{1}{4}\left( {\gamma  - \ln 2} \right) \lambda^2  + o({\lambda ^{5/2}}), &    d=6,  \\ 
 \frac{1}{{\rho _c^{(d)}}}\lambda  - {\omega _d}{\lambda ^2} + o({\lambda ^{5/2}}), & d\geq 7 ,
\end{array}  \right.
\end{eqnarray}
\end{widetext}
where $\gamma=0.5772...$ is the Euler-Mascheroni constant, and $\omega _d$ is a dimensionality-dependent parameter, given by
\begin{eqnarray}\label{eq5.5}
{\omega _d} = \frac{1}{{{d^2}\left( {d + 2} \right)}} + \frac{1}{{{{\left( {d - 4} \right)}^2}\left( {d - 6} \right)}}.
\end{eqnarray}
Furthermore, we expand $\lambda$ at $\lambda \sim 0$ as $\lambda=\lambda_1+\lambda_2+\cdots $, where 
\begin{eqnarray}\label{eq5.6}
\lambda_1=\rho_c^{(d)} \delta
\end{eqnarray}
is the linear part of the SCGF, and $\lambda_2$ is the part of the second-lowest order, which is a higher-order infinitesimal of $\delta$, i.e., $\lambda_2=o(\delta)$. Substituting this expansion into Eq.(\ref{eq5.4}), we obtain
\begin{eqnarray}\label{eq5.7}
\lambda_2 = \left\{ \begin{array}{llc}
(\rho_c^{(d)})^{5/2}{\delta ^{3/2}}, &     d=5,  \\
-\frac{1}{8}  (\rho_c^{(d)})^3 {\delta ^2}\ln ( {{\rho _c^{(d)}}\delta } ), &    d=6,  \\ 
 (\rho_c^{(d)})^3 {\omega _d}{\delta ^2}, & d\geq 7 .
\end{array}  \right.
\end{eqnarray}
Finally, by the Legendre-Fenchel transformation for $\lambda(k)$, we obtain the asymptotic forms of the rate function for $0<\rho-\rho_c \ll 1$, 
\begin{eqnarray}\label{eq5.8}
I(\rho) \simeq \left\{ \begin{array}{llc}
3\rho  + \frac{{1152}}{{3125}}{( {\rho  - {\rho _c^{(d)}}} )^3}, &     d=5,  \\
4\rho  + \frac{2 {( {\rho  - {\rho _c}} )^2}}{{{{( {{\rho _c^{(d)}}} )}^3}}}{[ { - \ln ( {{\rho _c^{(d)}}( {\rho  - {\rho _c^{(d)}}} )} )} ]^{ - 1}}, &    d=6,  \\ 
{k_c^{(d)}}\rho  + \frac{1}{{4(\rho _c^{(d)})^3{\omega_d}}} {{{( {\rho  - {\rho _c^{(d)}}} )}^2}}, & d\geq 7 . 
\end{array}  \right. \nonumber \\
\end{eqnarray}

In Fig.\ref{figd5}, we plot the SCGFs $\lambda(k)$ and rate functions for $d=5$ and $d=6$. As predicted, the SCGF $\lambda(k)$ given by quantum solution crosses zero at a nonzero $k_c^{(d)}=d-2$, marked by solid circles in Fig.\ref{figd5}(a). The first derivative of $\lambda(k)$ at $k_c^{(d)}$ is not continuous, leading to the singularity of the rate functions $I(\rho)$ at $\rho=\rho_c^{(d)}$. 
In the subcritical regime $0<\rho<\rho_c^{(d)}$, $I(\rho)$ is described by a linear branch $I(\rho)=k_c^{(d)} \rho=(d-2) \rho$. In the supercritical regime $\rho>\rho_c^{(d)}$, $I(\rho)$ is no longer a linear function of $\rho$, and contains higher terms of $\rho$, see the asymptotic results of Eq.(\ref{eq5.8}) for $\rho \gtrsim \rho_c^{(d)}$.

As demonstrated in previous studies \cite{nyawo2017minimal,nyawo2018dynamical,kanazawa2025universality,kanazawa2024dynamical}, the presence of a linear branch in $I(\rho)$ gives rise to temporal phase separation in dynamical trajectories. Specifically, a trajectory splits into two segments: the first is localized around $r=r_c$, and the second is nonlocalized. This phenomenon is analogous to spatial phase separation in thermodynamics, where the Helmholtz free energy depends linearly on particle density. The localized phase is atypical unless the value of $\rho$ is conditioned. 

In Fig.\ref{figsp}, we plot 100 samples of the radial trajectories of Brownian motion for $d=5$, where the value of $\rho$ is conditioned such that $\rho<\rho_c^{(5)}$ (a), $\rho \approx \rho_c^{(5)}$ (b), and $\rho>\rho_c^{(5)}$ (c). In simulations, the total duration is $T=5$, each time step is $\Delta=0.05$, and a cut-off $\epsilon=0.1$ is used to measure local time density (see Sec.\ref{sec_simu} for details of the simulations). To obtain 100 successful samples for a given $\rho$, we generate approximately  $1.2\times 10^6$, $4 \times 10^7$, and $2\times 10^9$ Brownian trajectories for the conditioned values of $\rho$: 0.7 (a), 0.85 (b), and 1.2 (c), respectively. For $\rho<\rho_c^{(5)}$, the qualitative trait of temporal phase separation between localized and nonlocalized states is evident. For $\rho \approx \rho_c^{(5)}$ and $\rho>\rho_c^{(5)}$, the localized state persists for a longer period. Nevertheless, full localization across the entire time domain—predicted theoretically in the limit $T \to \infty$—is not observed in our simulations, owing to the finiteness of $T$ in numerical experiments.

\section{Simulation verification}\label{sec_simu}
To simulate $d$-dimensional Brownian motions, the total time $T$ is divided into $K$ steps with each time step $\Delta t=T/K$. Thus, Eq.(\ref{eq1.1}) is discretized as
\begin{eqnarray}\label{eq6.1}
x_i^k = x_i^k + \sqrt {2D\Delta t} \eta _i^k,
\end{eqnarray}
where $x_i^k:=x_i(t=k\Delta t)$ denotes the $i$th component of the Brownian particle's position at time $t=k\Delta t$, and ${\eta}=\left\lbrace \eta _i^k \right\rbrace$ is a set of independent Gaussian random numbers with zero mean and unit variance with $i=1,\cdots,d$ and $k=1,\cdots,K$. These Gaussian random numbers are generated using the Box-Muller algorithm \cite{box1958note}. To obtain a trajectory of Brownian motions with duration $T$, we need to generate $Kd$ independent Gaussian random numbers.

\begin{figure*}
	\centerline{\includegraphics[width=1.8\columnwidth]{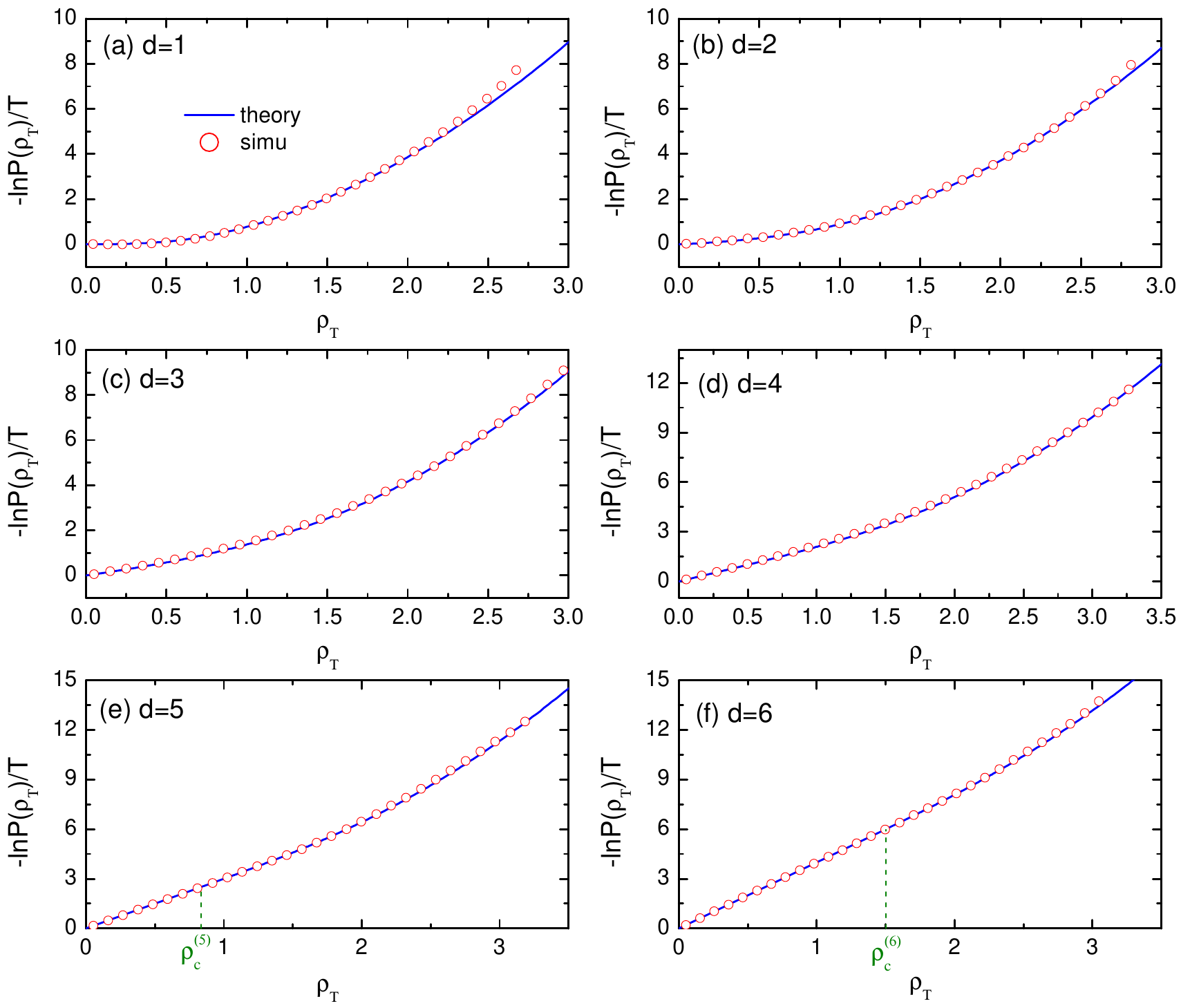}}
	\caption{Simulation verifications for the rate functions for $d=1,\cdots,6$ (from (a) to (f), respectively). Simulation results are indicated by symbols, and the rate functions predicted by our theory are shown by lines. In simulations, we set $T=10$, $\Delta t=0.01$ and $\epsilon=0.05$.     \label{figsimu}}
\end{figure*}

In order to compute the local time density $\rho_T$ at $r=r_c$ as defined in Eq.(\ref{eq1.2}), we use a cut-off $\epsilon$ to measure the fraction of time  spent in a domain $\left[  r_c-\epsilon,  r_c+\epsilon \right] $, and then divided by the domain width $2 \epsilon$. The mathematical definition is as follows. 
\begin{eqnarray}\label{eq6.2}
{\rho _T} = \frac{1}{T}\mathop {\lim }\limits_{\epsilon  \to 0} \frac{{{T_{2\epsilon }}( {{r_c}} )}}{{2\epsilon }}, 
\end{eqnarray}
with
\begin{eqnarray}\label{eq6.3}
{T_{2\epsilon }}( {{r_c}} ) = \int_0^T {{\mathds{1}_{\left[ {{r_c} - \epsilon ,{r_c} + \epsilon } \right]}}\left[  r (t)\right] } dt,
\end{eqnarray}
where $\mathds{1}_{\left[z_1, z_2 \right] }(z)$ is an indicator function defined as $\mathds{1}_{\left[z_1, z_2 \right] }(z)=1$ for $z \in \left[z_1, z_2 \right]$ and zero otherwise. In simulation, we choose a sufficiently small cut-off $\epsilon$ to get the probability density function of the local time density $P(\rho _T)$. 

By generating $n$ independent Brownian motion trajectories, we can construct the corresponding the histogram to estimate  the distribution of $\rho _T$. The distribution can be resolved down to a probability $o(1/n)$, e.g., $P \sim 10^{-6}$ for $10^6$ trajectories. However, for the tail of the distribution, i.e., the large deviation of $\rho _T$, such a direct simulation is prohibitively inefficient. To the end, we will employ a rare-event simulation approach, i.e., multiple histogram reweighting (MHR) \cite{ferrenberg1988new}, which is a statistical-mechanics inspired Monte Carlo (MC) method \cite{newman1999monte}. The method is computationally efficient and enables us to access the tail of $P(\rho _T)$. This approaches have been applied to many different problems, such as random graph properties \cite{hartmann2011large,schawe2019large,chen2021large}, random walks \cite{claussen2015convex,borjes2019large}, fractional Brownian motion \cite{hartmann2024first}, and the Kardar-Parisi-Zhang equation \cite{hartmann2018high,hartmann2019optimal,hartmann2021observing}.

To generate realizations of trajectories with extremely large values of $\rho _T$, we do not sample the random number according to its natural Gaussian product weight $G(\eta)$, but according to the modified weight $Q_{\beta}(\eta) \sim G(\eta) \exp[-\beta \rho _T(\eta) ]$, i.e., with an exponential bias. $\beta$ is an auxiliary inverse temperature. For $\beta=0$, the bias is absent, recovering to the original statistics. For $\beta<0$, it enables us to sample the distribution of $P(\rho _T)$ with large $\rho _T$.

To generate $\eta$ according to the modified weight $Q_{\beta}(\eta)$, we use a standard Markov-chain approach with the Metropolis-Hastings algorithm as follows \cite{newman1999monte}. First, we generate a random vector $\eta$ (containing $Kd$ independent Gaussian random numbers with zero mean and unit variance), and thus obtain an initial trajectory according to Eq.(\ref{eq6.1}) and the corresponding local time density around $r_c$, $\rho_T(\eta)$, according to Eqs.(\ref{eq6.2}) and (\ref{eq6.3}). Then, we generate a new random vector $\eta'$, where $\eta'$ is obtained by replacing one randomly selected random number in $\eta$ with a new random number. Surely, the new random number is also Gaussian distributed with zero mean and unit variance. We try to update the random vector (i.e., trajectory), $\eta \to \eta'$,  with a probability, $\min\left\lbrace 1, \exp(-\beta \Delta \rho_T)\right\rbrace $ , where $\Delta \rho_T=\rho_T(\eta')-\rho_T(\eta)$  is the difference of $\rho _T$ due the trial. Each MC step consists of $Kd$ trials of update. For a given inverse temperature $\beta_i$, the probability density $p_i(\rho _T)$ of generating a trajectory with the local time density $\rho _T$ follows the Boltzmann distribution,
\begin{eqnarray}\label{eq6.4}
p_i(\rho _T)=P(\rho _T)\frac{e^{-\beta_i \rho _T}}{Z_i},
\end{eqnarray}
where $P(\rho _T)$ is the probability density function of $\rho _T$ we want to obtain, and
$Z_i =\int d\rho _T P(\rho _T) e^{-\beta_i \rho _T}$ is the partition function (normalized factor) at the inverse temperature $\beta_i$. In practice, $p_i(\rho _T)$ can be obtained in simulations by collecting a histogram $N_i(\rho _T)$ of the number of times
out of $n_i$ that an interval $\left[\rho _T, \rho _T + d\rho _T \right) $ is observed, such that 
\begin{eqnarray}\label{eq6.5}
p_i(\rho _T) d\rho _T =\frac{N_i(\rho _T)}{n_i}.
\end{eqnarray}
Using Eq.(\ref{eq6.5}), Eq.(\ref{eq6.4}) can be rewritten as
\begin{eqnarray}\label{eq6.6}
P(\rho _T)d\rho _T=\frac{N_i(\rho _T) Z_i}{n_i e^{-\beta_i \rho _T}}.
\end{eqnarray}
The MHR takes advantage of collecting a series of histograms at nearby temperature overlap. We perform MC simulations for a series of different inverse temperatures $\beta_i$ chosen from an interval $\left[\beta_{\min} , \beta_{\max}\right]$. The improved estimate for $P(\rho _T)$ is given by
\begin{eqnarray}\label{eq6.7}
P(\rho _T)d\rho _T=\frac{\sum_{i} N_i(\rho _T)}{\sum_{j} n_j Z_j^{-1} e^{-\beta_j \rho _T}},
\end{eqnarray}
where the summations in the numerator and denominator are over all sampled inverse temperatures, and the partition function  $Z_j$ can be determined self-consistently by numerically solving the following equations,
\begin{eqnarray}\label{eq6.8}
Z_k =\int d\rho _T P(\rho _T) e^{-\beta_k \rho _T}=\int d\rho _T \frac{\sum_{i} N_i(\rho _T)}{\sum_{j} n_j Z_j^{-1} e^{(\beta_k-\beta_j) \rho _T}}. \nonumber \\
\end{eqnarray}

Simulation parameters are set as follows: total time $T=10$,  time step $\Delta {t}=0.01$, and cut-off $\epsilon=0.05$ for measuring the local time density. For MHR, we use 21 inverse temperatures uniformly spaced between $\beta_{\min}=-50$ and $\beta_{\max}=0$. 
For each $\beta_i$, we simulate $3 \times 10^4$ MC steps, in which 
the first $10^4$ MC steps are discarded to ensure equilibrium, and the remaining  $2 \times 10^4$ MC steps are used to accumulate  the histograms of $\rho _T$. To compare our theory results, we plot $-\ln{P(\rho _T)/T}$ (the minus of the logarithm of $P(\rho _T)$ scaled with $T$) versus  $\rho _T$. As shown by symbols in Fig.\ref{figsimu}, the simulation results for $d=1,\cdots,6$
are in excellent agreement with the theoretical rate function 
$I(\rho)$ (lines in Fig.\ref{figsimu}).

\section{Conclusions and Discussion}\label{sec5}
In conclusion, we have investigated the large-deviation statistics of the local time density $\rho_T$ of the radial component of a $d$-dimensional Brownian particle at a specific length. By solving the eigenvalue problem, we analytically derived  the SCGF $\lambda(k)$ of $\rho_T$ and correspondingly obtained the rate function $I(\rho)$ via the Legendre-Fenchel transform. For $d \leq 2$, both $\lambda(k)$ and $I(\rho)$ are analytic everywhere, indicating the absence of dynamical phase transitions. For $d>2$, $\lambda(k)$ exhibits a nonanalytic point at $k=k_c^{(d)}$, where $k_c^{(d)}=d-2$. Specifically, $\lambda(k)=0$ for $k<k_c^{(d)}$, as determined by the non-quantum eigenvalue, while $\lambda(k)>0$ for $k>k_c^{(d)}$, as given by the quantum eigenvalue. For $2<d\leq 4$, $\lambda(k)$ exhibits a second-order singularity at $k = k_c^{(d)}$: while its first derivative remains continuous across all $k$, the second derivative is discontinuous at this point. This distinctive singularity structure confirms the presence of a second-order DPT for $2 < d \leq 4$, and ensures the rate function $I(\rho)$ remains analytic everywhere. In contrast, for $d>4$, the first derivative of $\lambda(k)$ becomes discontinuous at $k_c^{(d)}$, rendering $I(\rho)$ nonanalytic. This gives rise to a singularity in the large-deviation rate function at $\rho=\rho_c^{(d)}$, where $\rho_c^{(d)}=d(d-4)/(2d-4)$—a signature of a first-order DPT. For $\rho<\rho_c^{(d)}$, $I(\rho)$ features a linear branch, whereas for $\rho>\rho_c^{(d)}$, $I(\rho)$ is a more complex function whose asymptotic behavior has been analyzed in detail. The transition occurs at $\rho=\rho_c^{(d)}$: above this value, Brownian trajectories are localized at the vicinity of the observed sphere shell throughout the entire dynamics; below this value, temporal phase separations of  dynamical trajectories emerges, where the particle resides near the observed sphere shell for a fraction of the time before escaping to infinity. Finally, we employ the MHR approach to sample the rare fluctuations of local time density, and numerically verify the DPT predicted by our theory.

In the context of Brownian motion, the recurrence-transience transition stands as a canonical example of a dimensionality-induced phase transition, wherein spatial dimension alone drives a qualitative shift in the system’s long-term dynamics \cite{klafter2011first}. Specifically, Brownian motion in one and two dimensions is recurrent: the particle returns to its starting point with probability one. In contrast, Brownian motion in three or higher dimensions is transient, with the particle having a probability less than one of revisiting its initial position. In the present work, we identify a new critical dimension, $d_c=4$ for the first-order DPT in the LDF of local time density, distinct from the critical dimension $d_c=2$ governing the recurrence-transience transition. Notably, the dimensionality-induced DPTs with temporal coexistence in our work shares qualitative similarities with critical behaviors in other stochastic systems. Specifically, it aligns with the dimension-dependent statistical transitions observed in the volume distribution of the Wiener sausage \cite{van2001moderate}, the number of distinct sites visited by random walkers \cite{phetpradap2011intersections}, and the area under a Bessel excursion \cite{kessler2014distribution}. These analogies highlight the importance of such a critical dimension in  governing fluctuation phenomena across stochastic processes.

The stochastic dynamics governing the radial component of high-dimensional Brownian motion is equivalent to one-dimensional diffusion within a logarithmic potential (i.e., the Bessel processes \cite{revuz2013continuous}). It is worth emphasizing that all results derived in this work are not restricted to integer dimensions $d$; they remain valid for non-integer $d$ as well. This is primarily because the modified Bessel functions (appeared in our derivations) are well-defined for non-integer orders. This logarithmic potential manifests in real-world systems—for instance, as an entropic term in the free energy cost associated with unzipping DNA base pairs to form denaturation bubbles \cite{bar2007loop,fogedby2007dna}, and in modeling the momentum spreading of cold atoms confined in optical lattices \cite{kessler2010infinite,PhysRevX.4.021036,aghion2017large}. Consequently, observing the DPTs described herein in actual low-dimensional systems is a promising task.

We note that our framework for the local-time density could also be extended to noninteracting multiparticle systems, leveraging the dimensional equivalence principle: For $N$ independent Brownian particles in $d$ dimensions, the problem effectively maps to a single particle in $Nd$ dimensions—analogous to the decomposition of $d$-dimensional single-particle motion into noninteracting one-dimensional motions \cite{kanazawa2025universality}. Such an extension, mirroring the multiparticle DPT scenarios enabled by $Nd >4 $, represents a promising direction to explore collective local-time behaviors in high-dimensional many-body systems.

A theoretical perspective is in order. Stochastic resetting describes a renewal process wherein dynamics are stochastically interrupted and subsequently restarted from the initial state. This topic has recently received considerable attention in both theoretical \cite{evans2020stochastic,evans2014diffusion,ray2020diffusion} and experimental \cite{tal2020experimental,faisant2021optimal} investigations. In high-dimensional spaces, however, several open questions remain: whether DPTs under stochastic resetting—acting on partial or all dimensions—can occur, and if they do, whether the critical dimension would be altered.

\appendix
\section{Derivation of the second equality of Eq.(\ref{eq2.5})}\label{appa}
Taking the derivative of Eq.(\ref{eq2.4}) with respect to $r$ and evaluating this result in the limit $r \to 1$, we obtain
\begin{widetext}
\begin{eqnarray}\label{eqa1.1}
{\phi' _k}( r ) = \left\{ \begin{array}{llc}
\frac{1}{2}{A_k}{\lambda ^{ - d/4}}\left[ {\lambda {I_{d/2 - 2}}( {\sqrt \lambda  } ) - \left( {d - 2} \right)\sqrt \lambda  {I_{d/2 - 1}}( {\sqrt \lambda  } ) + \lambda {I_{d/2}}( {\sqrt \lambda  } )} \right], &     r \to 1^-,  \\
 - \frac{1}{2}{B_k}{\lambda ^{\frac{{2 - d}}{4}}}\left[ {\sqrt \lambda  {K_{d/2 - 2}}( {\sqrt \lambda  } ) + \left( {d - 2} \right){K_{d/2 - 1}}( {\sqrt \lambda  } ) + \sqrt \lambda  {K_{d/2}}( {\sqrt \lambda  } )} \right], &    r \to 1^+,  \\ 
\end{array}  \right.
\end{eqnarray}
\end{widetext}
where we have used the derivative relations for modified Bessel functions: ${{I'}_\nu }( z ) = \frac{1}{2}{I_{\nu- 1}}( z ) + \frac{1}{2}{I_{\nu  + 1}}( z )$ and ${{K'}_\nu }( z ) =  - \frac{1}{2}{K_{\nu  - 1}}( z ) - \frac{1}{2}{K_{\nu  + 1}}( z )$ \cite{arfken2011mathematical}.

Using the recursive relations of modified Bessel functions \cite{arfken2011mathematical}, 
\begin{eqnarray}\label{eqa1.2}
\left\{ \begin{gathered}
{I_{\nu  - 1}}( z ) - {I_{\nu  + 1}}( z ) = \frac{{2\nu }}{z}{I_\nu }( z ) \hfill \\
{K_{\nu  - 1}}( z ) - {K_{\nu  + 1}}( z ) =  - \frac{{2\nu }}{z}{K_\nu }( z ) \hfill \\ 
\end{gathered}  \right.
\end{eqnarray}
Eq.(\ref{eqa1.1}) simplifies to
\begin{eqnarray}\label{eqa1.3}
{\phi _k}'( r ) = \left\{ \begin{array}{llc}
{A_k}{\lambda ^{1 - d/4}}{I_{d/2}}( {\sqrt \lambda  } ), &     r \to 1^-,  \\
 - {B_k}{\lambda ^{1 - d/4}}{K_{d/2}}( {\sqrt \lambda  } ), &    r \to 1^+.  \\ 
\end{array}  \right.
\end{eqnarray}
Furthermore, from Eq.(\ref{eq2.4}) we have ${\phi _k}( 1 ) = {A_k}{\lambda ^{1/2 - d/4}}{I_{d/2 - 1}}( {\sqrt \lambda  } )$. Substituting Eq.(\ref{eqa1.3}) and ${\phi _k}( 1 )$ into the matching condition, $\phi'_k(1^+)-\phi'_k(1^-)=-k \phi_k(1)$, yields the second equality in Eq.(\ref{eq2.5}).

\section{Asymptotic Behaviors of $\lambda(k)$ and $I(\rho)$ at large arguments}\label{appb}
This appendix supplements the asymptotic analyses of the SCGF $\lambda(k)$ (for large $k$) and the rate function $I(\rho)$ (for large $\rho$). The derivations below rely on the asymptotic properties of modified Bessel functions, which appear in eigenvalue equation (\ref{eq2.6}). At large $k$ (equivalently, large $\lambda$), the modified Bessel functions admit asymptotic expansions: $I_{\nu}(z) \sim \frac{e^z}{\sqrt{2 \pi z}} (1-\frac{4 \nu^2-1}{8z})$ and $K_{\nu}(z) \sim \sqrt{\frac{\pi}{2z}} e^{-z} (1+\frac{4 \nu^2-1}{8z})$. Notably, the terms containing $\nu$ (and thus dependent on the spatial dimension $d$) are subleading: as $z$ increases, their contribution becomes negligible compared to the leading exponential and inverse-square-root terms. To derive $\lambda(k)$’s large-$k$ behavior, we substitute the leading-order terms of $I_{\nu}(z)$ and $K_{\nu}(z)$ (neglecting subleading $\nu$-dependent terms) into the eigenvalue condition for $\lambda(k)$ (Eq. (\ref{eq2.6}) in the main text). This substitution eliminates dimension-dependent contributions, leading to a dimension-independent asymptotic result:  
\begin{eqnarray}\label{eqb1.1}
\lambda(k) \sim \frac{k^2}{4}, \quad k \to \infty.
\end{eqnarray}
The rate function $I(\rho)$ is linked to $\lambda(k)$ via the Legendre-Fenchel transform: $I(\rho) = \sup_{k \in \mathbb{R}} \left\{ k\rho - \lambda(k) \right\}$. For large $\rho$, the supremum is determined by the large-$k$ regime of $\lambda(k)$, where $\lambda(k) \sim \frac{k^2}{4}$ (from Eq. (\ref{eqb1.1})). To find the optimal $k^*$ that maximizes $k\rho - \frac{k^2}{4}$, we take the derivative with respect to $k$ and set it to zero, yielding $k^*=2\rho$. Substituting $k^* = 2\rho$ back into the transform gives
\begin{eqnarray}\label{eqb1.2}
I(\rho) \sim (2\rho )\rho -\frac{(2\rho)^2}{4}=\rho^2, \quad  \rho \to \infty.
\end{eqnarray}
This quadratic scaling of $I(\rho)$ is a hallmark of Gaussian-like fluctuations.

\begin{acknowledgments}
This work is supported by the National Natural Science Foundation of China (11875069), the Key Scientific Research Fund of Anhui Provincial Education Department (2023AH050116) and Anhui Project (Grant No. 2022AH020009).
\end{acknowledgments}


\begin{thebibliography}{98}
	\expandafter\ifx\csname natexlab\endcsname\relax\def\natexlab#1{#1}\fi
	\expandafter\ifx\csname bibnamefont\endcsname\relax
	\def\bibnamefont#1{#1}\fi
	\expandafter\ifx\csname bibfnamefont\endcsname\relax
	\def\bibfnamefont#1{#1}\fi
	\expandafter\ifx\csname citenamefont\endcsname\relax
	\def\citenamefont#1{#1}\fi
	\expandafter\ifx\csname url\endcsname\relax
	\def\url#1{\texttt{#1}}\fi
	\expandafter\ifx\csname urlprefix\endcsname\relax\def\urlprefix{URL }\fi
	\providecommand{\bibinfo}[2]{#2}
	\providecommand{\eprint}[2][]{\url{#2}}
	
	\bibitem[{\citenamefont{Derrida}(2007)}]{derrida2007non}
	\bibinfo{author}{\bibfnamefont{B.}~\bibnamefont{Derrida}},
	\bibinfo{journal}{Journal of Statistical Mechanics: Theory and Experiment}
	\textbf{\bibinfo{volume}{2007}}, \bibinfo{pages}{P07023}
	(\bibinfo{year}{2007}).
	
	\bibitem[{\citenamefont{Bertini et~al.}(2015)\citenamefont{Bertini, De~Sole,
			Gabrielli, Jona-Lasinio, and Landim}}]{bertini2015macroscopic}
	\bibinfo{author}{\bibfnamefont{L.}~\bibnamefont{Bertini}},
	\bibinfo{author}{\bibfnamefont{A.}~\bibnamefont{De~Sole}},
	\bibinfo{author}{\bibfnamefont{D.}~\bibnamefont{Gabrielli}},
	\bibinfo{author}{\bibfnamefont{G.}~\bibnamefont{Jona-Lasinio}},
	\bibnamefont{and} \bibinfo{author}{\bibfnamefont{C.}~\bibnamefont{Landim}},
	\bibinfo{journal}{Reviews of Modern Physics} \textbf{\bibinfo{volume}{87}},
	\bibinfo{pages}{593} (\bibinfo{year}{2015}).
	
	\bibitem[{\citenamefont{Cohen and Mukamel}(2012)}]{cohen2012phase}
	\bibinfo{author}{\bibfnamefont{O.}~\bibnamefont{Cohen}} \bibnamefont{and}
	\bibinfo{author}{\bibfnamefont{D.}~\bibnamefont{Mukamel}},
	\bibinfo{journal}{Physical Review Letters} \textbf{\bibinfo{volume}{108}},
	\bibinfo{pages}{060602} (\bibinfo{year}{2012}).
	
	\bibitem[{\citenamefont{Bodineau and Derrida}(2004)}]{bodineau2004current}
	\bibinfo{author}{\bibfnamefont{T.}~\bibnamefont{Bodineau}} \bibnamefont{and}
	\bibinfo{author}{\bibfnamefont{B.}~\bibnamefont{Derrida}},
	\bibinfo{journal}{Physical Review Letters} \textbf{\bibinfo{volume}{92}},
	\bibinfo{pages}{180601} (\bibinfo{year}{2004}).
	
	\bibitem[{\citenamefont{Harris et~al.}(2005)\citenamefont{Harris, R{\'a}kos,
			and Sch{\"u}tz}}]{harris2005current}
	\bibinfo{author}{\bibfnamefont{R.}~\bibnamefont{Harris}},
	\bibinfo{author}{\bibfnamefont{A.}~\bibnamefont{R{\'a}kos}},
	\bibnamefont{and} \bibinfo{author}{\bibfnamefont{G.~M.}
		\bibnamefont{Sch{\"u}tz}}, \bibinfo{journal}{Journal of Statistical
		Mechanics: Theory and Experiment} \textbf{\bibinfo{volume}{2005}},
	\bibinfo{pages}{P08003} (\bibinfo{year}{2005}).
	
	\bibitem[{\citenamefont{Ovaskainen and
			Meerson}(2010)}]{ovaskainen2010stochastic}
	\bibinfo{author}{\bibfnamefont{O.}~\bibnamefont{Ovaskainen}} \bibnamefont{and}
	\bibinfo{author}{\bibfnamefont{B.}~\bibnamefont{Meerson}},
	\bibinfo{journal}{Trends in ecology \& evolution}
	\textbf{\bibinfo{volume}{25}}, \bibinfo{pages}{643} (\bibinfo{year}{2010}).
	
	\bibitem[{\citenamefont{Assaf and Meerson}(2017)}]{assaf2017wkb}
	\bibinfo{author}{\bibfnamefont{M.}~\bibnamefont{Assaf}} \bibnamefont{and}
	\bibinfo{author}{\bibfnamefont{B.}~\bibnamefont{Meerson}},
	\bibinfo{journal}{Journal of Physics A: Mathematical and Theoretical}
	\textbf{\bibinfo{volume}{50}}, \bibinfo{pages}{263001}
	(\bibinfo{year}{2017}).
	
	\bibitem[{\citenamefont{Merhav and Kafri}(2010)}]{merhav2010bose}
	\bibinfo{author}{\bibfnamefont{N.}~\bibnamefont{Merhav}} \bibnamefont{and}
	\bibinfo{author}{\bibfnamefont{Y.}~\bibnamefont{Kafri}},
	\bibinfo{journal}{Journal of Statistical Mechanics: Theory and Experiment}
	\textbf{\bibinfo{volume}{2010}}, \bibinfo{pages}{P02011}
	(\bibinfo{year}{2010}).
	
	\bibitem[{\citenamefont{Szavits-Nossan
			et~al.}(2014)\citenamefont{Szavits-Nossan, Evans, and
			Majumdar}}]{szavits2014constraint}
	\bibinfo{author}{\bibfnamefont{J.}~\bibnamefont{Szavits-Nossan}},
	\bibinfo{author}{\bibfnamefont{M.~R.} \bibnamefont{Evans}}, \bibnamefont{and}
	\bibinfo{author}{\bibfnamefont{S.~N.} \bibnamefont{Majumdar}},
	\bibinfo{journal}{Physical Review Letters} \textbf{\bibinfo{volume}{112}},
	\bibinfo{pages}{020602} (\bibinfo{year}{2014}).
	
	\bibitem[{\citenamefont{Smith and Majumdar}(2022)}]{smith2022condensation}
	\bibinfo{author}{\bibfnamefont{N.~R.} \bibnamefont{Smith}} \bibnamefont{and}
	\bibinfo{author}{\bibfnamefont{S.~N.} \bibnamefont{Majumdar}},
	\bibinfo{journal}{Journal of Statistical Mechanics: Theory and Experiment}
	\textbf{\bibinfo{volume}{2022}}, \bibinfo{pages}{053212}
	(\bibinfo{year}{2022}).
	
	\bibitem[{\citenamefont{Seifert}(2012)}]{seifert2012stochastic}
	\bibinfo{author}{\bibfnamefont{U.}~\bibnamefont{Seifert}},
	\bibinfo{journal}{Reports on Progress in Physics}
	\textbf{\bibinfo{volume}{75}}, \bibinfo{pages}{126001}
	(\bibinfo{year}{2012}).
	
	\bibitem[{\citenamefont{Barato and Seifert}(2015)}]{barato2015thermodynamic}
	\bibinfo{author}{\bibfnamefont{A.~C.} \bibnamefont{Barato}} \bibnamefont{and}
	\bibinfo{author}{\bibfnamefont{U.}~\bibnamefont{Seifert}},
	\bibinfo{journal}{Physical Review Letters} \textbf{\bibinfo{volume}{114}},
	\bibinfo{pages}{158101} (\bibinfo{year}{2015}).
	
	\bibitem[{\citenamefont{Horowitz and
			Gingrich}(2020)}]{horowitz2020thermodynamic}
	\bibinfo{author}{\bibfnamefont{J.~M.} \bibnamefont{Horowitz}} \bibnamefont{and}
	\bibinfo{author}{\bibfnamefont{T.~R.} \bibnamefont{Gingrich}},
	\bibinfo{journal}{Nature Physics} \textbf{\bibinfo{volume}{16}},
	\bibinfo{pages}{15} (\bibinfo{year}{2020}).
	
	\bibitem[{\citenamefont{Smith}(2022)}]{smith2022anomalous}
	\bibinfo{author}{\bibfnamefont{N.~R.} \bibnamefont{Smith}},
	\bibinfo{journal}{Physical Review E} \textbf{\bibinfo{volume}{105}},
	\bibinfo{pages}{014120} (\bibinfo{year}{2022}).
	
	\bibitem[{\citenamefont{Nickelsen and Touchette}(2022)}]{nickelsen2022noise}
	\bibinfo{author}{\bibfnamefont{D.}~\bibnamefont{Nickelsen}} \bibnamefont{and}
	\bibinfo{author}{\bibfnamefont{H.}~\bibnamefont{Touchette}},
	\bibinfo{journal}{Physical Review E} \textbf{\bibinfo{volume}{105}},
	\bibinfo{pages}{064102} (\bibinfo{year}{2022}).
	
	\bibitem[{\citenamefont{Meerson}(2019)}]{meerson2019anomalous}
	\bibinfo{author}{\bibfnamefont{B.}~\bibnamefont{Meerson}},
	\bibinfo{journal}{Physical Review E} \textbf{\bibinfo{volume}{100}},
	\bibinfo{pages}{042135} (\bibinfo{year}{2019}).
	
	\bibitem[{\citenamefont{Ellis}(2012)}]{ellis2012entropy}
	\bibinfo{author}{\bibfnamefont{R.~S.} \bibnamefont{Ellis}},
	\emph{\bibinfo{title}{Entropy, large deviations, and statistical mechanics}},
	vol. \bibinfo{volume}{271} (\bibinfo{publisher}{Springer Science \& Business
		Media}, \bibinfo{year}{2012}).
	
	\bibitem[{\citenamefont{Oono}(1989)}]{oono1989large}
	\bibinfo{author}{\bibfnamefont{Y.}~\bibnamefont{Oono}},
	\bibinfo{journal}{Progress of Theoretical Physics Supplement}
	\textbf{\bibinfo{volume}{99}}, \bibinfo{pages}{165} (\bibinfo{year}{1989}).
	
	\bibitem[{\citenamefont{Hollander}(2000)}]{hollander2000large}
	\bibinfo{author}{\bibfnamefont{F.}~\bibnamefont{Hollander}},
	\emph{\bibinfo{title}{Large deviations}}, vol.~\bibinfo{volume}{14}
	(\bibinfo{publisher}{American Mathematical Soc.}, \bibinfo{year}{2000}).
	
	\bibitem[{\citenamefont{Jack}(2020)}]{jack2020ergodicity}
	\bibinfo{author}{\bibfnamefont{R.~L.} \bibnamefont{Jack}},
	\bibinfo{journal}{The European Physical Journal B}
	\textbf{\bibinfo{volume}{93}}, \bibinfo{pages}{74} (\bibinfo{year}{2020}).
	
	\bibitem[{\citenamefont{Touchette}(2009)}]{touchette2009large}
	\bibinfo{author}{\bibfnamefont{H.}~\bibnamefont{Touchette}},
	\bibinfo{journal}{Physics Reports} \textbf{\bibinfo{volume}{478}},
	\bibinfo{pages}{1} (\bibinfo{year}{2009}).
	
	\bibitem[{\citenamefont{Bodineau and Derrida}(2005)}]{bodineau2005distribution}
	\bibinfo{author}{\bibfnamefont{T.}~\bibnamefont{Bodineau}} \bibnamefont{and}
	\bibinfo{author}{\bibfnamefont{B.}~\bibnamefont{Derrida}},
	\bibinfo{journal}{Physical Review E} \textbf{\bibinfo{volume}{72}},
	\bibinfo{pages}{066110} (\bibinfo{year}{2005}).
	
	\bibitem[{\citenamefont{Bertini et~al.}(2005)\citenamefont{Bertini, De~Sole,
			Gabrielli, Jona-Lasinio, and Landim}}]{bertini2005current}
	\bibinfo{author}{\bibfnamefont{L.}~\bibnamefont{Bertini}},
	\bibinfo{author}{\bibfnamefont{A.}~\bibnamefont{De~Sole}},
	\bibinfo{author}{\bibfnamefont{D.}~\bibnamefont{Gabrielli}},
	\bibinfo{author}{\bibfnamefont{G.}~\bibnamefont{Jona-Lasinio}},
	\bibnamefont{and} \bibinfo{author}{\bibfnamefont{C.}~\bibnamefont{Landim}},
	\bibinfo{journal}{Physical Review Letters} \textbf{\bibinfo{volume}{94}},
	\bibinfo{pages}{030601} (\bibinfo{year}{2005}).
	
	\bibitem[{\citenamefont{Bertini et~al.}(2006)\citenamefont{Bertini, Sole,
			Gabrielli, Jona-Lasinio, and Landim}}]{bertini2006non}
	\bibinfo{author}{\bibfnamefont{L.}~\bibnamefont{Bertini}},
	\bibinfo{author}{\bibfnamefont{A.~D.} \bibnamefont{Sole}},
	\bibinfo{author}{\bibfnamefont{D.}~\bibnamefont{Gabrielli}},
	\bibinfo{author}{\bibfnamefont{G.}~\bibnamefont{Jona-Lasinio}},
	\bibnamefont{and} \bibinfo{author}{\bibfnamefont{C.}~\bibnamefont{Landim}},
	\bibinfo{journal}{Journal of Statistical Physics}
	\textbf{\bibinfo{volume}{123}}, \bibinfo{pages}{237} (\bibinfo{year}{2006}).
	
	\bibitem[{\citenamefont{Agranov et~al.}(2023)\citenamefont{Agranov, Cates, and
			Jack}}]{agranov2023tricritical}
	\bibinfo{author}{\bibfnamefont{T.}~\bibnamefont{Agranov}},
	\bibinfo{author}{\bibfnamefont{M.~E.} \bibnamefont{Cates}}, \bibnamefont{and}
	\bibinfo{author}{\bibfnamefont{R.~L.} \bibnamefont{Jack}},
	\bibinfo{journal}{Physical Review Letters} \textbf{\bibinfo{volume}{131}},
	\bibinfo{pages}{017102} (\bibinfo{year}{2023}).
	
	\bibitem[{\citenamefont{Baek and Kafri}(2015)}]{baek2015singularities}
	\bibinfo{author}{\bibfnamefont{Y.}~\bibnamefont{Baek}} \bibnamefont{and}
	\bibinfo{author}{\bibfnamefont{Y.}~\bibnamefont{Kafri}},
	\bibinfo{journal}{Journal of Statistical Mechanics: Theory and Experiment}
	\textbf{\bibinfo{volume}{2015}}, \bibinfo{pages}{P08026}
	(\bibinfo{year}{2015}).
	
	\bibitem[{\citenamefont{Baek et~al.}(2017)\citenamefont{Baek, Kafri, and
			Lecomte}}]{baek2017dynamical}
	\bibinfo{author}{\bibfnamefont{Y.}~\bibnamefont{Baek}},
	\bibinfo{author}{\bibfnamefont{Y.}~\bibnamefont{Kafri}}, \bibnamefont{and}
	\bibinfo{author}{\bibfnamefont{V.}~\bibnamefont{Lecomte}},
	\bibinfo{journal}{Physical Review Letters} \textbf{\bibinfo{volume}{118}},
	\bibinfo{pages}{030604} (\bibinfo{year}{2017}).
	
	\bibitem[{\citenamefont{Bertini et~al.}(2010)\citenamefont{Bertini, De~Sole,
			Gabrielli, Jona-Lasinio, and Landim}}]{bertini2010lagrangian}
	\bibinfo{author}{\bibfnamefont{L.}~\bibnamefont{Bertini}},
	\bibinfo{author}{\bibfnamefont{A.}~\bibnamefont{De~Sole}},
	\bibinfo{author}{\bibfnamefont{D.}~\bibnamefont{Gabrielli}},
	\bibinfo{author}{\bibfnamefont{G.}~\bibnamefont{Jona-Lasinio}},
	\bibnamefont{and} \bibinfo{author}{\bibfnamefont{C.}~\bibnamefont{Landim}},
	\bibinfo{journal}{Journal of Statistical Mechanics: Theory and Experiment}
	\textbf{\bibinfo{volume}{2010}}, \bibinfo{pages}{L11001}
	(\bibinfo{year}{2010}).
	
	\bibitem[{\citenamefont{Bunin et~al.}(2013)\citenamefont{Bunin, Kafri, and
			Podolsky}}]{bunin2013cusp}
	\bibinfo{author}{\bibfnamefont{G.}~\bibnamefont{Bunin}},
	\bibinfo{author}{\bibfnamefont{Y.}~\bibnamefont{Kafri}}, \bibnamefont{and}
	\bibinfo{author}{\bibfnamefont{D.}~\bibnamefont{Podolsky}},
	\bibinfo{journal}{Journal of Statistical Physics}
	\textbf{\bibinfo{volume}{152}}, \bibinfo{pages}{112} (\bibinfo{year}{2013}).
	
	\bibitem[{\citenamefont{Espigares et~al.}(2013)\citenamefont{Espigares,
			Garrido, and Hurtado}}]{espigares2013dynamical}
	\bibinfo{author}{\bibfnamefont{C.~P.} \bibnamefont{Espigares}},
	\bibinfo{author}{\bibfnamefont{P.~L.} \bibnamefont{Garrido}},
	\bibnamefont{and} \bibinfo{author}{\bibfnamefont{P.~I.}
		\bibnamefont{Hurtado}}, \bibinfo{journal}{Physical Review E}
	\textbf{\bibinfo{volume}{87}}, \bibinfo{pages}{032115}
	(\bibinfo{year}{2013}).
	
	\bibitem[{\citenamefont{Aminov et~al.}(2014)\citenamefont{Aminov, Bunin, and
			Kafri}}]{aminov2014singularities}
	\bibinfo{author}{\bibfnamefont{A.}~\bibnamefont{Aminov}},
	\bibinfo{author}{\bibfnamefont{G.}~\bibnamefont{Bunin}}, \bibnamefont{and}
	\bibinfo{author}{\bibfnamefont{Y.}~\bibnamefont{Kafri}},
	\bibinfo{journal}{Journal of Statistical Mechanics: Theory and Experiment}
	\textbf{\bibinfo{volume}{2014}}, \bibinfo{pages}{P08017}
	(\bibinfo{year}{2014}).
	
	\bibitem[{\citenamefont{Shpielberg et~al.}(2017)\citenamefont{Shpielberg, Don,
			and Akkermans}}]{shpielberg2017numerical}
	\bibinfo{author}{\bibfnamefont{O.}~\bibnamefont{Shpielberg}},
	\bibinfo{author}{\bibfnamefont{Y.}~\bibnamefont{Don}}, \bibnamefont{and}
	\bibinfo{author}{\bibfnamefont{E.}~\bibnamefont{Akkermans}},
	\bibinfo{journal}{Physical Review E} \textbf{\bibinfo{volume}{95}},
	\bibinfo{pages}{032137} (\bibinfo{year}{2017}).
	
	\bibitem[{\citenamefont{Garrahan et~al.}(2007)\citenamefont{Garrahan, Jack,
			Lecomte, Pitard, van Duijvendijk, and van Wijland}}]{garrahan2007dynamical}
	\bibinfo{author}{\bibfnamefont{J.~P.} \bibnamefont{Garrahan}},
	\bibinfo{author}{\bibfnamefont{R.~L.} \bibnamefont{Jack}},
	\bibinfo{author}{\bibfnamefont{V.}~\bibnamefont{Lecomte}},
	\bibinfo{author}{\bibfnamefont{E.}~\bibnamefont{Pitard}},
	\bibinfo{author}{\bibfnamefont{K.}~\bibnamefont{van Duijvendijk}},
	\bibnamefont{and} \bibinfo{author}{\bibfnamefont{F.}~\bibnamefont{van
			Wijland}}, \bibinfo{journal}{Physical Review Letters}
	\textbf{\bibinfo{volume}{98}}, \bibinfo{pages}{195702}
	(\bibinfo{year}{2007}).
	
	\bibitem[{\citenamefont{Garrahan et~al.}(2009)\citenamefont{Garrahan, Jack,
			Lecomte, Pitard, van Duijvendijk, and van Wijland}}]{garrahan2009first}
	\bibinfo{author}{\bibfnamefont{J.~P.} \bibnamefont{Garrahan}},
	\bibinfo{author}{\bibfnamefont{R.~L.} \bibnamefont{Jack}},
	\bibinfo{author}{\bibfnamefont{V.}~\bibnamefont{Lecomte}},
	\bibinfo{author}{\bibfnamefont{E.}~\bibnamefont{Pitard}},
	\bibinfo{author}{\bibfnamefont{K.}~\bibnamefont{van Duijvendijk}},
	\bibnamefont{and} \bibinfo{author}{\bibfnamefont{F.}~\bibnamefont{van
			Wijland}}, \bibinfo{journal}{Journal of Physics A: Mathematical and
		Theoretical} \textbf{\bibinfo{volume}{42}}, \bibinfo{pages}{075007}
	(\bibinfo{year}{2009}).
	
	\bibitem[{\citenamefont{Kumar et~al.}(2011)\citenamefont{Kumar, Ramaswamy, and
			Sood}}]{kumar2011symmetry}
	\bibinfo{author}{\bibfnamefont{N.}~\bibnamefont{Kumar}},
	\bibinfo{author}{\bibfnamefont{S.}~\bibnamefont{Ramaswamy}},
	\bibnamefont{and} \bibinfo{author}{\bibfnamefont{A.}~\bibnamefont{Sood}},
	\bibinfo{journal}{Physical Review Letters} \textbf{\bibinfo{volume}{106}},
	\bibinfo{pages}{118001} (\bibinfo{year}{2011}).
	
	\bibitem[{\citenamefont{Cagnetta et~al.}(2017)\citenamefont{Cagnetta, Corberi,
			Gonnella, and Suma}}]{cagnetta2017large}
	\bibinfo{author}{\bibfnamefont{F.}~\bibnamefont{Cagnetta}},
	\bibinfo{author}{\bibfnamefont{F.}~\bibnamefont{Corberi}},
	\bibinfo{author}{\bibfnamefont{G.}~\bibnamefont{Gonnella}}, \bibnamefont{and}
	\bibinfo{author}{\bibfnamefont{A.}~\bibnamefont{Suma}},
	\bibinfo{journal}{Physical Review Letters} \textbf{\bibinfo{volume}{119}},
	\bibinfo{pages}{158002} (\bibinfo{year}{2017}).
	
	\bibitem[{\citenamefont{Nemoto et~al.}(2019)\citenamefont{Nemoto, Fodor, Cates,
			Jack, and Tailleur}}]{nemoto2019optimizing}
	\bibinfo{author}{\bibfnamefont{T.}~\bibnamefont{Nemoto}},
	\bibinfo{author}{\bibfnamefont{{\'E}.}~\bibnamefont{Fodor}},
	\bibinfo{author}{\bibfnamefont{M.~E.} \bibnamefont{Cates}},
	\bibinfo{author}{\bibfnamefont{R.~L.} \bibnamefont{Jack}}, \bibnamefont{and}
	\bibinfo{author}{\bibfnamefont{J.}~\bibnamefont{Tailleur}},
	\bibinfo{journal}{Physical Review E} \textbf{\bibinfo{volume}{99}},
	\bibinfo{pages}{022605} (\bibinfo{year}{2019}).
	
	\bibitem[{\citenamefont{Semeraro et~al.}(2023)\citenamefont{Semeraro, Gonnella,
			Suma, and Zamparo}}]{semeraro2023work}
	\bibinfo{author}{\bibfnamefont{M.}~\bibnamefont{Semeraro}},
	\bibinfo{author}{\bibfnamefont{G.}~\bibnamefont{Gonnella}},
	\bibinfo{author}{\bibfnamefont{A.}~\bibnamefont{Suma}}, \bibnamefont{and}
	\bibinfo{author}{\bibfnamefont{M.}~\bibnamefont{Zamparo}},
	\bibinfo{journal}{Physical Review Letters} \textbf{\bibinfo{volume}{131}},
	\bibinfo{pages}{158302} (\bibinfo{year}{2023}).
	
	\bibitem[{\citenamefont{De~Bacco et~al.}(2016)\citenamefont{De~Bacco, Guggiola,
			K{\"u}hn, and Paga}}]{de2016rare}
	\bibinfo{author}{\bibfnamefont{C.}~\bibnamefont{De~Bacco}},
	\bibinfo{author}{\bibfnamefont{A.}~\bibnamefont{Guggiola}},
	\bibinfo{author}{\bibfnamefont{R.}~\bibnamefont{K{\"u}hn}}, \bibnamefont{and}
	\bibinfo{author}{\bibfnamefont{P.}~\bibnamefont{Paga}},
	\bibinfo{journal}{Journal of Physics A: Mathematical and Theoretical}
	\textbf{\bibinfo{volume}{49}}, \bibinfo{pages}{184003}
	(\bibinfo{year}{2016}).
	
	\bibitem[{\citenamefont{Coghi et~al.}(2019)\citenamefont{Coghi, Morand, and
			Touchette}}]{coghi2019large}
	\bibinfo{author}{\bibfnamefont{F.}~\bibnamefont{Coghi}},
	\bibinfo{author}{\bibfnamefont{J.}~\bibnamefont{Morand}}, \bibnamefont{and}
	\bibinfo{author}{\bibfnamefont{H.}~\bibnamefont{Touchette}},
	\bibinfo{journal}{Physical Review E} \textbf{\bibinfo{volume}{99}},
	\bibinfo{pages}{022137} (\bibinfo{year}{2019}).
	
	\bibitem[{\citenamefont{Carugno et~al.}(2023)\citenamefont{Carugno, Vivo, and
			Coghi}}]{carugno2023delocalization}
	\bibinfo{author}{\bibfnamefont{G.}~\bibnamefont{Carugno}},
	\bibinfo{author}{\bibfnamefont{P.}~\bibnamefont{Vivo}}, \bibnamefont{and}
	\bibinfo{author}{\bibfnamefont{F.}~\bibnamefont{Coghi}},
	\bibinfo{journal}{Physical Review E} \textbf{\bibinfo{volume}{107}},
	\bibinfo{pages}{024126} (\bibinfo{year}{2023}).
	
	\bibitem[{\citenamefont{Speck et~al.}(2012)\citenamefont{Speck, Engel, and
			Seifert}}]{speck2012large}
	\bibinfo{author}{\bibfnamefont{T.}~\bibnamefont{Speck}},
	\bibinfo{author}{\bibfnamefont{A.}~\bibnamefont{Engel}}, \bibnamefont{and}
	\bibinfo{author}{\bibfnamefont{U.}~\bibnamefont{Seifert}},
	\bibinfo{journal}{Journal of Statistical Mechanics: Theory and Experiment}
	\textbf{\bibinfo{volume}{2012}}, \bibinfo{pages}{P12001}
	(\bibinfo{year}{2012}).
	
	\bibitem[{\citenamefont{Tsobgni~Nyawo and Touchette}(2016)}]{tsobgni2016large}
	\bibinfo{author}{\bibfnamefont{P.}~\bibnamefont{Tsobgni~Nyawo}}
	\bibnamefont{and}
	\bibinfo{author}{\bibfnamefont{H.}~\bibnamefont{Touchette}},
	\bibinfo{journal}{Physical Review E} \textbf{\bibinfo{volume}{94}},
	\bibinfo{pages}{032101} (\bibinfo{year}{2016}).
	
	\bibitem[{\citenamefont{Meerson and Smith}(2019)}]{meerson2019geometrical}
	\bibinfo{author}{\bibfnamefont{B.}~\bibnamefont{Meerson}} \bibnamefont{and}
	\bibinfo{author}{\bibfnamefont{N.~R.} \bibnamefont{Smith}},
	\bibinfo{journal}{Journal of Physics A: Mathematical and Theoretical}
	\textbf{\bibinfo{volume}{52}}, \bibinfo{pages}{415001}
	(\bibinfo{year}{2019}).
	
	\bibitem[{\citenamefont{Mukherjee and
			Smith}(2025)}]{mukherjee2025nonequilibrium}
	\bibinfo{author}{\bibfnamefont{S.}~\bibnamefont{Mukherjee}} \bibnamefont{and}
	\bibinfo{author}{\bibfnamefont{N.~R.} \bibnamefont{Smith}},
	\bibinfo{journal}{Journal of Statistical Mechanics: Theory and Experiment}
	\textbf{\bibinfo{volume}{2025}}, \bibinfo{pages}{033205}
	(\bibinfo{year}{2025}).
	
	\bibitem[{\citenamefont{Majumdar et~al.}(2015)\citenamefont{Majumdar,
			Sabhapandit, and Schehr}}]{majumdar2015dynamical}
	\bibinfo{author}{\bibfnamefont{S.~N.} \bibnamefont{Majumdar}},
	\bibinfo{author}{\bibfnamefont{S.}~\bibnamefont{Sabhapandit}},
	\bibnamefont{and} \bibinfo{author}{\bibfnamefont{G.}~\bibnamefont{Schehr}},
	\bibinfo{journal}{Physical Review E} \textbf{\bibinfo{volume}{91}},
	\bibinfo{pages}{052131} (\bibinfo{year}{2015}).
	
	\bibitem[{\citenamefont{Harris and Touchette}(2017)}]{harris2017phase}
	\bibinfo{author}{\bibfnamefont{R.~J.} \bibnamefont{Harris}} \bibnamefont{and}
	\bibinfo{author}{\bibfnamefont{H.}~\bibnamefont{Touchette}},
	\bibinfo{journal}{Journal of Physics A: Mathematical and Theoretical}
	\textbf{\bibinfo{volume}{50}}, \bibinfo{pages}{10LT01}
	(\bibinfo{year}{2017}).
	
	\bibitem[{\citenamefont{Nyawo and Touchette}(2017)}]{nyawo2017minimal}
	\bibinfo{author}{\bibfnamefont{P.~T.} \bibnamefont{Nyawo}} \bibnamefont{and}
	\bibinfo{author}{\bibfnamefont{H.}~\bibnamefont{Touchette}},
	\bibinfo{journal}{Europhysics Letters} \textbf{\bibinfo{volume}{116}},
	\bibinfo{pages}{50009} (\bibinfo{year}{2017}).
	
	\bibitem[{\citenamefont{Nyawo and Touchette}(2018)}]{nyawo2018dynamical}
	\bibinfo{author}{\bibfnamefont{P.~T.} \bibnamefont{Nyawo}} \bibnamefont{and}
	\bibinfo{author}{\bibfnamefont{H.}~\bibnamefont{Touchette}},
	\bibinfo{journal}{Physical Review E} \textbf{\bibinfo{volume}{98}},
	\bibinfo{pages}{052103} (\bibinfo{year}{2018}).
	
	\bibitem[{\citenamefont{Mukherjee et~al.}(2024)\citenamefont{Mukherjee,
			Le~Doussal, and Smith}}]{mukherjee2024large}
	\bibinfo{author}{\bibfnamefont{S.}~\bibnamefont{Mukherjee}},
	\bibinfo{author}{\bibfnamefont{P.}~\bibnamefont{Le~Doussal}},
	\bibnamefont{and} \bibinfo{author}{\bibfnamefont{N.~R.} \bibnamefont{Smith}},
	\bibinfo{journal}{Physical Review E} \textbf{\bibinfo{volume}{110}},
	\bibinfo{pages}{024107} (\bibinfo{year}{2024}).
	
	\bibitem[{\citenamefont{Mukherjee and Smith}(2023)}]{mukherjee2023dynamical}
	\bibinfo{author}{\bibfnamefont{S.}~\bibnamefont{Mukherjee}} \bibnamefont{and}
	\bibinfo{author}{\bibfnamefont{N.~R.} \bibnamefont{Smith}},
	\bibinfo{journal}{Physical Review E} \textbf{\bibinfo{volume}{107}},
	\bibinfo{pages}{064133} (\bibinfo{year}{2023}).
	
	\bibitem[{\citenamefont{Yerrababu et~al.}(2024)\citenamefont{Yerrababu,
			Majumdar, and Sadhu}}]{yerrababu2024dynamical}
	\bibinfo{author}{\bibfnamefont{Y.~R.} \bibnamefont{Yerrababu}},
	\bibinfo{author}{\bibfnamefont{S.~N.} \bibnamefont{Majumdar}},
	\bibnamefont{and} \bibinfo{author}{\bibfnamefont{T.}~\bibnamefont{Sadhu}},
	\bibinfo{journal}{arXiv preprint arXiv:2412.19516}  (\bibinfo{year}{2024}).
	
	\bibitem[{\citenamefont{Gu{\'e}neau et~al.}(2025)\citenamefont{Gu{\'e}neau,
			Majumdar, and Schehr}}]{gueneau2025large}
	\bibinfo{author}{\bibfnamefont{M.}~\bibnamefont{Gu{\'e}neau}},
	\bibinfo{author}{\bibfnamefont{S.~N.} \bibnamefont{Majumdar}},
	\bibnamefont{and} \bibinfo{author}{\bibfnamefont{G.}~\bibnamefont{Schehr}},
	\bibinfo{journal}{Physical Review Letters} \textbf{\bibinfo{volume}{135}},
	\bibinfo{pages}{067102} (\bibinfo{year}{2025}).
	
	\bibitem[{\citenamefont{Kanazawa et~al.}(2025)\citenamefont{Kanazawa,
			Kawaguchi, and Adachi}}]{kanazawa2025universality}
	\bibinfo{author}{\bibfnamefont{T.}~\bibnamefont{Kanazawa}},
	\bibinfo{author}{\bibfnamefont{K.}~\bibnamefont{Kawaguchi}},
	\bibnamefont{and} \bibinfo{author}{\bibfnamefont{K.}~\bibnamefont{Adachi}},
	\bibinfo{journal}{Physical Review Research} \textbf{\bibinfo{volume}{7}},
	\bibinfo{pages}{013124} (\bibinfo{year}{2025}).
	
	\bibitem[{\citenamefont{Kanazawa et~al.}(2024)\citenamefont{Kanazawa,
			Kawaguchi, and Adachi}}]{kanazawa2024dynamical}
	\bibinfo{author}{\bibfnamefont{T.}~\bibnamefont{Kanazawa}},
	\bibinfo{author}{\bibfnamefont{K.}~\bibnamefont{Kawaguchi}},
	\bibnamefont{and} \bibinfo{author}{\bibfnamefont{K.}~\bibnamefont{Adachi}},
	\bibinfo{journal}{arXiv preprint arXiv:2407.18282}  (\bibinfo{year}{2024}).
	
	\bibitem[{\citenamefont{Wilemski and Fixman}(1973)}]{wilemski1973general}
	\bibinfo{author}{\bibfnamefont{G.}~\bibnamefont{Wilemski}} \bibnamefont{and}
	\bibinfo{author}{\bibfnamefont{M.}~\bibnamefont{Fixman}},
	\bibinfo{journal}{The Journal of Chemical Physics}
	\textbf{\bibinfo{volume}{58}}, \bibinfo{pages}{4009} (\bibinfo{year}{1973}).
	
	\bibitem[{\citenamefont{Doi}(1975)}]{doi1975theory}
	\bibinfo{author}{\bibfnamefont{M.}~\bibnamefont{Doi}},
	\bibinfo{journal}{Chemical Physics} \textbf{\bibinfo{volume}{11}},
	\bibinfo{pages}{115} (\bibinfo{year}{1975}).
	
	\bibitem[{\citenamefont{B{\'e}nichou and Voituriez}(2014)}]{benichou2014first}
	\bibinfo{author}{\bibfnamefont{O.}~\bibnamefont{B{\'e}nichou}}
	\bibnamefont{and}
	\bibinfo{author}{\bibfnamefont{R.}~\bibnamefont{Voituriez}},
	\bibinfo{journal}{Physics Reports} \textbf{\bibinfo{volume}{539}},
	\bibinfo{pages}{225} (\bibinfo{year}{2014}).
	
	\bibitem[{\citenamefont{Koshland}(1980)}]{koshland1980bacterial}
	\bibinfo{author}{\bibfnamefont{D.}~\bibnamefont{Koshland}},
	\emph{\bibinfo{title}{Bacterial chemotaxis as a model behavioral system}}
	(\bibinfo{publisher}{Raven, New York}, \bibinfo{year}{1980}).
	
	\bibitem[{\citenamefont{Kishore and Kundu}(2021)}]{kishore2021local}
	\bibinfo{author}{\bibfnamefont{G.}~\bibnamefont{Kishore}} \bibnamefont{and}
	\bibinfo{author}{\bibfnamefont{A.}~\bibnamefont{Kundu}},
	\bibinfo{journal}{Journal of Statistical Mechanics: Theory and Experiment}
	\textbf{\bibinfo{volume}{2021}}, \bibinfo{pages}{033218}
	(\bibinfo{year}{2021}).
	
	\bibitem[{\citenamefont{Majumdar and Comtet}(2002)}]{majumdar2002local}
	\bibinfo{author}{\bibfnamefont{S.~N.} \bibnamefont{Majumdar}} \bibnamefont{and}
	\bibinfo{author}{\bibfnamefont{A.}~\bibnamefont{Comtet}},
	\bibinfo{journal}{Physical Review Letters} \textbf{\bibinfo{volume}{89}},
	\bibinfo{pages}{060601} (\bibinfo{year}{2002}).
	
	\bibitem[{\citenamefont{Sabhapandit et~al.}(2006)\citenamefont{Sabhapandit,
			Majumdar, and Comtet}}]{sabhapandit2006statistical}
	\bibinfo{author}{\bibfnamefont{S.}~\bibnamefont{Sabhapandit}},
	\bibinfo{author}{\bibfnamefont{S.~N.} \bibnamefont{Majumdar}},
	\bibnamefont{and} \bibinfo{author}{\bibfnamefont{A.}~\bibnamefont{Comtet}},
	\bibinfo{journal}{Physical Review E} \textbf{\bibinfo{volume}{73}},
	\bibinfo{pages}{051102} (\bibinfo{year}{2006}).
	
	\bibitem[{\citenamefont{Pal et~al.}(2019)\citenamefont{Pal, Chatterjee,
			Reuveni, and Kundu}}]{pal2019local}
	\bibinfo{author}{\bibfnamefont{A.}~\bibnamefont{Pal}},
	\bibinfo{author}{\bibfnamefont{R.}~\bibnamefont{Chatterjee}},
	\bibinfo{author}{\bibfnamefont{S.}~\bibnamefont{Reuveni}}, \bibnamefont{and}
	\bibinfo{author}{\bibfnamefont{A.}~\bibnamefont{Kundu}},
	\bibinfo{journal}{Journal of Physics A: Mathematical and Theoretical}
	\textbf{\bibinfo{volume}{52}}, \bibinfo{pages}{264002}
	(\bibinfo{year}{2019}).
	
	\bibitem[{\citenamefont{Singh and Pal}(2022)}]{singh2022first}
	\bibinfo{author}{\bibfnamefont{P.}~\bibnamefont{Singh}} \bibnamefont{and}
	\bibinfo{author}{\bibfnamefont{A.}~\bibnamefont{Pal}},
	\bibinfo{journal}{Journal of Physics A: Mathematical and Theoretical}
	\textbf{\bibinfo{volume}{55}}, \bibinfo{pages}{234001}
	(\bibinfo{year}{2022}).
	
	\bibitem[{\citenamefont{Comtet et~al.}(2002)\citenamefont{Comtet, Desbois, and
			Majumdar}}]{comtet2002local}
	\bibinfo{author}{\bibfnamefont{A.}~\bibnamefont{Comtet}},
	\bibinfo{author}{\bibfnamefont{J.}~\bibnamefont{Desbois}}, \bibnamefont{and}
	\bibinfo{author}{\bibfnamefont{S.~N.} \bibnamefont{Majumdar}},
	\bibinfo{journal}{Journal of Physics A: Mathematical and General}
	\textbf{\bibinfo{volume}{35}}, \bibinfo{pages}{L687} (\bibinfo{year}{2002}).
	
	\bibitem[{\citenamefont{Singh and Kundu}(2021)}]{singh2021local}
	\bibinfo{author}{\bibfnamefont{P.}~\bibnamefont{Singh}} \bibnamefont{and}
	\bibinfo{author}{\bibfnamefont{A.}~\bibnamefont{Kundu}},
	\bibinfo{journal}{Physical Review E} \textbf{\bibinfo{volume}{103}},
	\bibinfo{pages}{042119} (\bibinfo{year}{2021}).
	
	\bibitem[{\citenamefont{Burenev et~al.}(2023)\citenamefont{Burenev, Majumdar,
			and Rosso}}]{burenev2023local}
	\bibinfo{author}{\bibfnamefont{I.~N.} \bibnamefont{Burenev}},
	\bibinfo{author}{\bibfnamefont{S.~N.} \bibnamefont{Majumdar}},
	\bibnamefont{and} \bibinfo{author}{\bibfnamefont{A.}~\bibnamefont{Rosso}},
	\bibinfo{journal}{Physical Review E} \textbf{\bibinfo{volume}{108}},
	\bibinfo{pages}{064113} (\bibinfo{year}{2023}).
	
	\bibitem[{\citenamefont{Smith and Meerson}(2024)}]{smith2024macroscopic}
	\bibinfo{author}{\bibfnamefont{N.~R.} \bibnamefont{Smith}} \bibnamefont{and}
	\bibinfo{author}{\bibfnamefont{B.}~\bibnamefont{Meerson}},
	\bibinfo{journal}{Physica A: Statistical Mechanics and its Applications}
	\textbf{\bibinfo{volume}{639}}, \bibinfo{pages}{129616}
	(\bibinfo{year}{2024}).
	
	\bibitem[{\citenamefont{Touchette}(2018)}]{touchette2018introduction}
	\bibinfo{author}{\bibfnamefont{H.}~\bibnamefont{Touchette}},
	\bibinfo{journal}{Physica A: Statistical Mechanics and its Applications}
	\textbf{\bibinfo{volume}{504}}, \bibinfo{pages}{5} (\bibinfo{year}{2018}).
	
	\bibitem[{\citenamefont{Arfken et~al.}(2011)\citenamefont{Arfken, Weber, and
			Harris}}]{arfken2011mathematical}
	\bibinfo{author}{\bibfnamefont{G.~B.} \bibnamefont{Arfken}},
	\bibinfo{author}{\bibfnamefont{H.~J.} \bibnamefont{Weber}}, \bibnamefont{and}
	\bibinfo{author}{\bibfnamefont{F.~E.} \bibnamefont{Harris}},
	\emph{\bibinfo{title}{Mathematical methods for physicists: a comprehensive
			guide}} (\bibinfo{publisher}{Academic press}, \bibinfo{year}{2011}).
	
	\bibitem[{\citenamefont{Touchette}(2005)}]{touchette2005legendre}
	\bibinfo{author}{\bibfnamefont{H.}~\bibnamefont{Touchette}},
	\bibinfo{journal}{URL
		https://ise.ncsu.edu/wp-content/uploads/sites/9/2015/07/or706-LF-transform-1.pdf}
	(\bibinfo{year}{2005}).
	
	\bibitem[{\citenamefont{Box and Muller}(1958)}]{box1958note}
	\bibinfo{author}{\bibfnamefont{G.~E.} \bibnamefont{Box}} \bibnamefont{and}
	\bibinfo{author}{\bibfnamefont{M.~E.} \bibnamefont{Muller}},
	\bibinfo{journal}{The Annals of Mathematical Statistics}
	\textbf{\bibinfo{volume}{29}}, \bibinfo{pages}{610} (\bibinfo{year}{1958}).
	
	\bibitem[{\citenamefont{Ferrenberg and Swendsen}(1988)}]{ferrenberg1988new}
	\bibinfo{author}{\bibfnamefont{A.~M.} \bibnamefont{Ferrenberg}}
	\bibnamefont{and} \bibinfo{author}{\bibfnamefont{R.~H.}
		\bibnamefont{Swendsen}}, \bibinfo{journal}{Physical Review Letters}
	\textbf{\bibinfo{volume}{61}}, \bibinfo{pages}{2635} (\bibinfo{year}{1988}).
	
	\bibitem[{\citenamefont{Newman and Barkema}(1999)}]{newman1999monte}
	\bibinfo{author}{\bibfnamefont{M.~E.} \bibnamefont{Newman}} \bibnamefont{and}
	\bibinfo{author}{\bibfnamefont{G.~T.} \bibnamefont{Barkema}},
	\emph{\bibinfo{title}{Monte Carlo methods in statistical physics}}
	(\bibinfo{publisher}{Clarendon Press}, \bibinfo{year}{1999}).
	
	\bibitem[{\citenamefont{Hartmann}(2011)}]{hartmann2011large}
	\bibinfo{author}{\bibfnamefont{A.~K.} \bibnamefont{Hartmann}},
	\bibinfo{journal}{The European Physical Journal B}
	\textbf{\bibinfo{volume}{84}}, \bibinfo{pages}{627} (\bibinfo{year}{2011}).
	
	\bibitem[{\citenamefont{Schawe and Hartmann}(2019)}]{schawe2019large}
	\bibinfo{author}{\bibfnamefont{H.}~\bibnamefont{Schawe}} \bibnamefont{and}
	\bibinfo{author}{\bibfnamefont{A.~K.} \bibnamefont{Hartmann}},
	\bibinfo{journal}{The European Physical Journal B}
	\textbf{\bibinfo{volume}{92}}, \bibinfo{pages}{73} (\bibinfo{year}{2019}).
	
	\bibitem[{\citenamefont{Chen et~al.}(2021)\citenamefont{Chen, Huang, Shen, Li,
			and Zhang}}]{chen2021large}
	\bibinfo{author}{\bibfnamefont{H.}~\bibnamefont{Chen}},
	\bibinfo{author}{\bibfnamefont{F.}~\bibnamefont{Huang}},
	\bibinfo{author}{\bibfnamefont{C.}~\bibnamefont{Shen}},
	\bibinfo{author}{\bibfnamefont{G.}~\bibnamefont{Li}}, \bibnamefont{and}
	\bibinfo{author}{\bibfnamefont{H.}~\bibnamefont{Zhang}},
	\bibinfo{journal}{Journal of Statistical Mechanics: Theory and Experiment}
	\textbf{\bibinfo{volume}{2021}}, \bibinfo{pages}{113402}
	(\bibinfo{year}{2021}).
	
	\bibitem[{\citenamefont{Claussen et~al.}(2015)\citenamefont{Claussen, Hartmann,
			and Majumdar}}]{claussen2015convex}
	\bibinfo{author}{\bibfnamefont{G.}~\bibnamefont{Claussen}},
	\bibinfo{author}{\bibfnamefont{A.~K.} \bibnamefont{Hartmann}},
	\bibnamefont{and} \bibinfo{author}{\bibfnamefont{S.~N.}
		\bibnamefont{Majumdar}}, \bibinfo{journal}{Physical Review E}
	\textbf{\bibinfo{volume}{91}}, \bibinfo{pages}{052104}
	(\bibinfo{year}{2015}).
	
	\bibitem[{\citenamefont{B{\"o}rjes et~al.}(2019)\citenamefont{B{\"o}rjes,
			Schawe, and Hartmann}}]{borjes2019large}
	\bibinfo{author}{\bibfnamefont{J.}~\bibnamefont{B{\"o}rjes}},
	\bibinfo{author}{\bibfnamefont{H.}~\bibnamefont{Schawe}}, \bibnamefont{and}
	\bibinfo{author}{\bibfnamefont{A.~K.} \bibnamefont{Hartmann}},
	\bibinfo{journal}{Physical Review E} \textbf{\bibinfo{volume}{99}},
	\bibinfo{pages}{042104} (\bibinfo{year}{2019}).
	
	\bibitem[{\citenamefont{Hartmann and Meerson}(2024)}]{hartmann2024first}
	\bibinfo{author}{\bibfnamefont{A.~K.} \bibnamefont{Hartmann}} \bibnamefont{and}
	\bibinfo{author}{\bibfnamefont{B.}~\bibnamefont{Meerson}},
	\bibinfo{journal}{Physical Review E} \textbf{\bibinfo{volume}{109}},
	\bibinfo{pages}{014146} (\bibinfo{year}{2024}).
	
	\bibitem[{\citenamefont{Hartmann et~al.}(2018)\citenamefont{Hartmann,
			Le~Doussal, Majumdar, Rosso, and Schehr}}]{hartmann2018high}
	\bibinfo{author}{\bibfnamefont{A.~K.} \bibnamefont{Hartmann}},
	\bibinfo{author}{\bibfnamefont{P.}~\bibnamefont{Le~Doussal}},
	\bibinfo{author}{\bibfnamefont{S.~N.} \bibnamefont{Majumdar}},
	\bibinfo{author}{\bibfnamefont{A.}~\bibnamefont{Rosso}}, \bibnamefont{and}
	\bibinfo{author}{\bibfnamefont{G.}~\bibnamefont{Schehr}},
	\bibinfo{journal}{Europhysics Letters} \textbf{\bibinfo{volume}{121}},
	\bibinfo{pages}{67004} (\bibinfo{year}{2018}).
	
	\bibitem[{\citenamefont{Hartmann et~al.}(2019)\citenamefont{Hartmann, Meerson,
			and Sasorov}}]{hartmann2019optimal}
	\bibinfo{author}{\bibfnamefont{A.~K.} \bibnamefont{Hartmann}},
	\bibinfo{author}{\bibfnamefont{B.}~\bibnamefont{Meerson}}, \bibnamefont{and}
	\bibinfo{author}{\bibfnamefont{P.}~\bibnamefont{Sasorov}},
	\bibinfo{journal}{Physical Review Research} \textbf{\bibinfo{volume}{1}},
	\bibinfo{pages}{032043} (\bibinfo{year}{2019}).
	
	\bibitem[{\citenamefont{Hartmann et~al.}(2021)\citenamefont{Hartmann, Meerson,
			and Sasorov}}]{hartmann2021observing}
	\bibinfo{author}{\bibfnamefont{A.~K.} \bibnamefont{Hartmann}},
	\bibinfo{author}{\bibfnamefont{B.}~\bibnamefont{Meerson}}, \bibnamefont{and}
	\bibinfo{author}{\bibfnamefont{P.}~\bibnamefont{Sasorov}},
	\bibinfo{journal}{Physical Review E} \textbf{\bibinfo{volume}{104}},
	\bibinfo{pages}{054125} (\bibinfo{year}{2021}).
	
	\bibitem[{\citenamefont{Klafter and Sokolov}(2011)}]{klafter2011first}
	\bibinfo{author}{\bibfnamefont{J.}~\bibnamefont{Klafter}} \bibnamefont{and}
	\bibinfo{author}{\bibfnamefont{I.~M.} \bibnamefont{Sokolov}},
	\emph{\bibinfo{title}{First steps in random walks: from tools to
			applications}} (\bibinfo{publisher}{OUP Oxford}, \bibinfo{year}{2011}).
	
	\bibitem[{\citenamefont{van~den Berg et~al.}(2001)\citenamefont{van~den Berg,
			Bolthausen, and den Hollander}}]{van2001moderate}
	\bibinfo{author}{\bibfnamefont{M.}~\bibnamefont{van~den Berg}},
	\bibinfo{author}{\bibfnamefont{E.}~\bibnamefont{Bolthausen}},
	\bibnamefont{and} \bibinfo{author}{\bibfnamefont{F.}~\bibnamefont{den
			Hollander}}, \bibinfo{journal}{Annals of Mathematics}
	\textbf{\bibinfo{volume}{153}}, \bibinfo{pages}{355} (\bibinfo{year}{2001}).
	
	\bibitem[{\citenamefont{Phetpradap}(2011)}]{phetpradap2011intersections}
	\bibinfo{author}{\bibfnamefont{P.}~\bibnamefont{Phetpradap}}, Ph.D. thesis,
	\bibinfo{school}{University of Bath} (\bibinfo{year}{2011}).
	
	\bibitem[{\citenamefont{Kessler et~al.}(2014)\citenamefont{Kessler, Medalion,
			and Barkai}}]{kessler2014distribution}
	\bibinfo{author}{\bibfnamefont{D.~A.} \bibnamefont{Kessler}},
	\bibinfo{author}{\bibfnamefont{S.}~\bibnamefont{Medalion}}, \bibnamefont{and}
	\bibinfo{author}{\bibfnamefont{E.}~\bibnamefont{Barkai}},
	\bibinfo{journal}{Journal of Statistical Physics}
	\textbf{\bibinfo{volume}{156}}, \bibinfo{pages}{686} (\bibinfo{year}{2014}).
	
	\bibitem[{\citenamefont{Revuz and Yor}(2013)}]{revuz2013continuous}
	\bibinfo{author}{\bibfnamefont{D.}~\bibnamefont{Revuz}} \bibnamefont{and}
	\bibinfo{author}{\bibfnamefont{M.}~\bibnamefont{Yor}},
	\emph{\bibinfo{title}{Continuous martingales and Brownian motion}}, vol.
	\bibinfo{volume}{293} (\bibinfo{publisher}{Springer Science \& Business
		Media}, \bibinfo{year}{2013}).
	
	\bibitem[{\citenamefont{Bar et~al.}(2007)\citenamefont{Bar, Kafri, and
			Mukamel}}]{bar2007loop}
	\bibinfo{author}{\bibfnamefont{A.}~\bibnamefont{Bar}},
	\bibinfo{author}{\bibfnamefont{Y.}~\bibnamefont{Kafri}}, \bibnamefont{and}
	\bibinfo{author}{\bibfnamefont{D.}~\bibnamefont{Mukamel}},
	\bibinfo{journal}{Physical Review Letters} \textbf{\bibinfo{volume}{98}},
	\bibinfo{pages}{038103} (\bibinfo{year}{2007}).
	
	\bibitem[{\citenamefont{Fogedby and Metzler}(2007)}]{fogedby2007dna}
	\bibinfo{author}{\bibfnamefont{H.~C.} \bibnamefont{Fogedby}} \bibnamefont{and}
	\bibinfo{author}{\bibfnamefont{R.}~\bibnamefont{Metzler}},
	\bibinfo{journal}{Physical Review Letters} \textbf{\bibinfo{volume}{98}},
	\bibinfo{pages}{070601} (\bibinfo{year}{2007}).
	
	\bibitem[{\citenamefont{Kessler and Barkai}(2010)}]{kessler2010infinite}
	\bibinfo{author}{\bibfnamefont{D.~A.} \bibnamefont{Kessler}} \bibnamefont{and}
	\bibinfo{author}{\bibfnamefont{E.}~\bibnamefont{Barkai}},
	\bibinfo{journal}{Physical Review Letters} \textbf{\bibinfo{volume}{105}},
	\bibinfo{pages}{120602} (\bibinfo{year}{2010}).
	
	\bibitem[{\citenamefont{Barkai et~al.}(2014)\citenamefont{Barkai, Aghion, and
			Kessler}}]{PhysRevX.4.021036}
	\bibinfo{author}{\bibfnamefont{E.}~\bibnamefont{Barkai}},
	\bibinfo{author}{\bibfnamefont{E.}~\bibnamefont{Aghion}}, \bibnamefont{and}
	\bibinfo{author}{\bibfnamefont{D.~A.} \bibnamefont{Kessler}},
	\bibinfo{journal}{Physical Review X} \textbf{\bibinfo{volume}{4}},
	\bibinfo{pages}{021036} (\bibinfo{year}{2014}).
	
	\bibitem[{\citenamefont{Aghion et~al.}(2017)\citenamefont{Aghion, Kessler, and
			Barkai}}]{aghion2017large}
	\bibinfo{author}{\bibfnamefont{E.}~\bibnamefont{Aghion}},
	\bibinfo{author}{\bibfnamefont{D.~A.} \bibnamefont{Kessler}},
	\bibnamefont{and} \bibinfo{author}{\bibfnamefont{E.}~\bibnamefont{Barkai}},
	\bibinfo{journal}{Physical Review Letters} \textbf{\bibinfo{volume}{118}},
	\bibinfo{pages}{260601} (\bibinfo{year}{2017}).
	
	\bibitem[{\citenamefont{Evans et~al.}(2020)\citenamefont{Evans, Majumdar, and
			Schehr}}]{evans2020stochastic}
	\bibinfo{author}{\bibfnamefont{M.~R.} \bibnamefont{Evans}},
	\bibinfo{author}{\bibfnamefont{S.~N.} \bibnamefont{Majumdar}},
	\bibnamefont{and} \bibinfo{author}{\bibfnamefont{G.}~\bibnamefont{Schehr}},
	\bibinfo{journal}{Journal of Physics A: Mathematical and Theoretical}
	\textbf{\bibinfo{volume}{53}}, \bibinfo{pages}{193001}
	(\bibinfo{year}{2020}).
	
	\bibitem[{\citenamefont{Evans and Majumdar}(2014)}]{evans2014diffusion}
	\bibinfo{author}{\bibfnamefont{M.~R.} \bibnamefont{Evans}} \bibnamefont{and}
	\bibinfo{author}{\bibfnamefont{S.~N.} \bibnamefont{Majumdar}},
	\bibinfo{journal}{Journal of Physics A: Mathematical and Theoretical}
	\textbf{\bibinfo{volume}{47}}, \bibinfo{pages}{285001}
	(\bibinfo{year}{2014}).
	
	\bibitem[{\citenamefont{Ray and Reuveni}(2020)}]{ray2020diffusion}
	\bibinfo{author}{\bibfnamefont{S.}~\bibnamefont{Ray}} \bibnamefont{and}
	\bibinfo{author}{\bibfnamefont{S.}~\bibnamefont{Reuveni}},
	\bibinfo{journal}{The Journal of Chemical Physics}
	\textbf{\bibinfo{volume}{152}} (\bibinfo{year}{2020}).
	
	\bibitem[{\citenamefont{Tal-Friedman et~al.}(2020)\citenamefont{Tal-Friedman,
			Pal, Sekhon, Reuveni, and Roichman}}]{tal2020experimental}
	\bibinfo{author}{\bibfnamefont{O.}~\bibnamefont{Tal-Friedman}},
	\bibinfo{author}{\bibfnamefont{A.}~\bibnamefont{Pal}},
	\bibinfo{author}{\bibfnamefont{A.}~\bibnamefont{Sekhon}},
	\bibinfo{author}{\bibfnamefont{S.}~\bibnamefont{Reuveni}}, \bibnamefont{and}
	\bibinfo{author}{\bibfnamefont{Y.}~\bibnamefont{Roichman}},
	\bibinfo{journal}{The Journal of Physical Chemistry Letters}
	\textbf{\bibinfo{volume}{11}}, \bibinfo{pages}{7350} (\bibinfo{year}{2020}).
	
	\bibitem[{\citenamefont{Faisant et~al.}(2021)\citenamefont{Faisant, Besga,
			Petrosyan, Ciliberto, and Majumdar}}]{faisant2021optimal}
	\bibinfo{author}{\bibfnamefont{F.}~\bibnamefont{Faisant}},
	\bibinfo{author}{\bibfnamefont{B.}~\bibnamefont{Besga}},
	\bibinfo{author}{\bibfnamefont{A.}~\bibnamefont{Petrosyan}},
	\bibinfo{author}{\bibfnamefont{S.}~\bibnamefont{Ciliberto}},
	\bibnamefont{and} \bibinfo{author}{\bibfnamefont{S.~N.}
		\bibnamefont{Majumdar}}, \bibinfo{journal}{Journal of Statistical Mechanics:
		Theory and Experiment} \textbf{\bibinfo{volume}{2021}},
	\bibinfo{pages}{113203} (\bibinfo{year}{2021}).
	
\end{thebibliography}

\end{document}